\documentclass[]{interact}

\usepackage{epstopdf}
\usepackage[caption=false]{subfig}
\usepackage{multirow}
\usepackage{makecell}
\usepackage{booktabs}
    \usepackage[dvipsnames, svgnames, table]{xcolor} 
    \usepackage{array, cellspace, boldline, float}
\usepackage{adjustbox}
\usepackage{colortbl}

\usepackage[numbers,sort&compress]{natbib}
\bibpunct[, ]{[}{]}{,}{n}{,}{,}
\renewcommand\bibfont{\fontsize{10}{12}\selectfont}

\theoremstyle{plain}

\theoremstyle{definition}

\theoremstyle{remark}

\newcommand{\pd}[2]{\frac{\partial #1}{\partial #2}}

\newcommand{\de}[2]{\frac{d #1}{d #2}}

\begin{document}


\title{Global redistribution and local migration in semi-discrete host-parasitoid population dynamic models.}

\author{
\name{Brooks Emerick\textsuperscript{a}\thanks{CONTACT Brooks Emerick. Email: bemerick@kutztown.edu} and Abhyudai Singh\textsuperscript{b}}
\affil{\textsuperscript{a} Kutztown University, Kutztown, PA 15200 \textsuperscript{b} University of Delaware, Newark, DE 19711}
}

\maketitle

\begin{abstract}
Host-parasitoid population dynamics is often probed using a semi-discrete/hybrid modeling framework. Here, the update functions in the discrete-time model connecting year-to-year changes in the population densities are obtained by solving ordinary differential equations that mechanistically describe interactions when hosts become vulnerable to parasitoid attacks. We use this semi-discrete formalism to study two key spatial effects: local movement (migration) of parasitoids between patches during the vulnerable period; and yearly redistribution of populations across patches outside the vulnerable period.  Our results show that in the absence of any redistribution, constant density-independent migration and parasitoid attack rates are unable to stabilize an otherwise unstable host-parasitoid population dynamics. Interestingly, inclusion of host redistribution (but not parasitoid redistribution)  before the start of the vulnerable period can lead to stable coexistence of both species. Next, we consider a Type-III functional response (parasitoid attack rate increases with host density), where the absence of any spatial effects leads to a neutrally stable host-parasitoid equilibrium.  As before, density-independent  parasitoid  migration by itself  is again insufficient  to stabilize the population dynamics and host redistribution provides a stabilizing influence. Finally, we show that a Type-III functional response combined with density-dependent parasitoid migration leads to stable coexistence, even in the absence of population redistributions. In summary, we have systematically characterized  parameter regimes leading to stable/unstable population dynamics with different forms of spatial heterogeneity coupled to the parasitoid's functional response using mechanistically formulated semi-discrete models. 
\end{abstract}


\section{Introduction}\label{Sec1}

The typical host-parasitoid interaction involves an interval within a given year for which the parasitoid and host are in direct contact with each other.  This interval is known as the vulnerable period as host larvae are susceptible to oviposition by the female parasitoid.  Generally, a parasitoid species spends a proportion of the vulnerable period searching for a suitable host \cite{Waage_1986, Murdoch_2003}.  The landscape in which a female parasitoid forages may be spatially heterogeneous in nature, consisting a several host larvae patches.  Studies have shown that once a parasitoid is within a suitable patch, they may not leave.  In fact, they will search until reaching the boundary and turn around to continue searching \cite{Loke_1984, van_Alphen_2003, Outreman_2005}.  However, search efficiency is key for reproduction and parasitoids may cut their losses and flee to a different patch in search of a suitable host \cite{Waage_1986, Bukovinszky_2012, Godfray_1994}.  The decision for a parasitoid to stay or leave a particular patch is based on several key factors: the concentration of kairomones in the patch which corresponds directly to the density of current host population, the presence of competitors such as hyperparasitoids, the number of times the female has made a successful egg deposit, and past experiences in a particular patch \cite{Outreman_2005, van_Alphen_2003, Driessen_1999, Wajnberg_1999, Hemerik_1993, Haccou_1991}.  This decision making process as well as the transit time between patches may provide an inefficiency in the parasitoid's search tactics that yields persistence in the host and parasitoid interaction.  For a more in depth review of patch-decision making rules and time allocation, we refer the interested reader to \cite{Cronin_2005} and \cite{van_Alphen_2003}.

Theoretical dynamics of spatially distributed host-parasitoid systems have been extensively studied \cite{Hassell_2000, Hassell_2000b}.  The transfer of hosts and parasitoids among patches is considered a global form of migration if a redistribution of hosts and parasitoids occurs among all patches in each generation.  This type of migration is known to have stabilizing effects on the coexistence equilibrium of host-parasitoid systems \cite{Hassell_1973, Hassell_1974, Hassell_1991b, Lett_2003}.  Local migration is characterized by hosts and parasitoids dispersing only to adjacent patches within each generation.  This is also known to yield persistence of the host-parasitoid interaction \cite{Hassell_1991a, Taylor_1993, Comins_1992, Rohani_1995,Singh_2009} provided there are enough patches.  These discrete-time models have captured the conditions necessary for host-parasitoid persistence using classic parasitism dynamics within each patch such as Nicholson-Bailey \cite{Nicholson_1935} or May \cite{May_1978}.  Adler et.~al.~\cite{Adler_1993} show that relatively low migration rates result in the persistence stable oscillations whereas Reeve \cite{Reeve_1985} and Rohani \cite{Rohani_1995} report destabilizing effects of migration if the dispersal occurs within a patch.  Lett et.~al.~\cite{Lett_2003} consider a discrete-time two-patch model where global migration can occur more than once before parasitism at each patch takes effect.  They show that the frequency of migration has a stabilizing effect as well.  The aggregate model considered by Lett et.~al.~is similar to the present model in Section \ref{Sec333} when a global redistribution of hosts and parasitoids is applied to the semi-discrete model.  However, we present results that couple the global migration proportions of both hosts and parasitoids outside the vulnerable (i.e. redistribution) with local migration that occurs during the vulnerable period.  In short, the semi-discrete approach allows for a generalization of these past models.  

Parasitoid and host migration have also been coupled with other dynamic characteristics of the parasitoid species.  Huang et.~al.~\cite{Huang_2016} created a continuous model between two patches that compare effects of the host-parasitoid interaction with autoparasitism and migration between both patches.  They find that in the two-patch model, density-dependent migration does have a stabilizing effect in addition to the stabilizing nature of autoparasitism.  The authors derive a host density dependent cross-migration term that considers both searching time and handling time of a female parasitoid.  In the present model, we consider a similar density dependent migration and report that also has a stabilizing effect on the location migration scale.  Reigada et.~al.~\cite{Reigada_2012} consider migration coupled with forager interference and sex ratio control.  They find that sex ratio alone cannot stabilize the persistence of the host-parasitoid interaction but the role of competitive interference influences the capability of a parasitoid to reduce host populations.  On a more general approach, Ngoc et.~al.~\cite{Ngoc_2010} also employ a continuous model to measure the effects of local patch migration in competing species models.  In particular, they investigate the competition-colonization trade-off and report that fast asymmetric migration can lead to the inferior competitor dominating a particular homogeneous environment.  An extensive review is given by Briggs and Hoopes in \cite{Briggs_2004}.  In summary, the migration dynamic has been implemented into both discrete and continuous models alike, and the overall modeling approaches have elucidated the effects of the migratory behavior on the persistence of host-parasitoid populations.  We wish to investigate similar effects with a hybrid approach.

Our aim is to formulate a general two-patch migration model using the semi-discrete modeling framework \cite{Emerick_2016,Singh_2007}, and analyze the conditions for coexistence with respect to migration parameters.  As we have seen in past models, stability of the coexistence equilibrium point usually occurs when the parasitoid is inefficient, indicating that if there is an interaction that inhibits the parasitoid's ability to oviposit then coexistence is more likely to occur.  We investigate this migration induced inefficiency in two forms: global redistribution of populations not occurring within the vulnerable period and a local parasitoid migration between patches during the vulnerable period.  We assume that only female parasitoids migrate during the vulnerable period and the host larvae are immobile.  Within each patch, we consider two types of parasitism: constant parasitism (i.e. Nicholson-Bailey model) and functional response. The paper is organized as follows: we formulate the semi-discrete two-patch model in Section \ref{Sec2} and provide numerical results for constant migration and density-dependent migration with constant parasitism at each location; in Section \ref{Sec3}, we derive several models with constant migration under the assumption that one or both populations redistribute to each patch every generation; Section \ref{Sec4} consists of the same models with a functional response in parasitism; and we conclude with a discussion in Section \ref{Sec5}.


\section{Model Formulation}\label{Sec2}
We consider the population of female hosts and parasitoids at two locations.  The host and parasitoid populations at site $i$ in year $t$ are denoted $H_{i,t}$ and $P_{i,t}$, respectively.  We assume that the yearly update for hosts and parasitoids at each location is given by the following discrete system:
\begin{align}
H_{i,t+1} & = L_i(T,t)\label{H} \\
P_{i,t+1}& = kI_i(T,t), \label{P}\end{align}
where $L_i(T,t)$ is the number of host larvae escaping parasitism after the vulnerable period at location $i$ and $kI_i(T,t)$ is the number of infected host larvae at the end of the vulnerable period at location $i$, where $k$ is the average number of parasitoid larvae that emerge from one infected host.  We define the host larvae, infected host larvae, and parasitoid populations as $L_i(\tau,t)$, $I_i(\tau,t)$, and $P_i(\tau, t)$, respectively.  We will assume that at each location basic parasitism occurs via the parasitic attack rate $g_i(\, \cdot \,)$ for $i = 1, 2$.  Here, we write `$(\, \cdot \,)$' to mean that this function may be density dependent.  Similarly, since the parasitoids are mobile and the host larvae are not, we assume that the parasitoids can migrate from location to location via the migration rates $g_{ij}(\, \cdot\,)$, defined as the potentially density dependent rate from location $i$ to location $j$.  The reactions can be written in terms of the following differential equations:

\begin{minipage}{.475\textwidth}
\begin{align}
\de{L_1}{\tau} & =- g_{1}(\, \cdot \,) L_1 P_1 \label{dL1_dt}\\
\de{I_1}{\tau} & = g_{1}(\, \cdot \,) L_1 P_1 \\
\de{P_1}{\tau} & =- g_{12}(\, \cdot \,)P_1 + g_{21}(\, \cdot \,) P_2\end{align}
\end{minipage}
\begin{minipage}{.475\textwidth}
\begin{align}
\de{L_2}{\tau} & =- g_{2}(\, \cdot \,) L_2 P_2 \\
\de{I_2}{\tau} & = g_{2}(\, \cdot \,) L_2 P_2 \\
\de{P_2}{\tau} & = g_{12}(\, \cdot \,)P_1 - g_{21}(\, \cdot \,) P_2, \label{dP2_dt}\end{align}
\end{minipage}\vspace{.55cm}
		
\noindent where $\tau \in [0,T]$ represents the time variable over the vulnerable period.  Here, we subject each population above to the following initial conditions:

\begin{minipage}{.475\textwidth}
\begin{align}
L_1(0,t) & = RH_{1,t} \label{ic_start} \\ 
I_1(0,t) & = 0 \\
P_1(0,t) & = P_{1,t} \label{P1_ic} \end{align} \end{minipage}
\begin{minipage}{.475\textwidth}
\begin{align}
L_2(0,t) & = RH_{2,t} \label{L2_ic} \\ 
I_2(0,t) & = 0 \\
P_2(0,t) & = P_{2,t}, \label{ic_end}\end{align} 
\end{minipage}\vspace{.55cm}

\noindent where $R>1$ denotes the number of viable eggs per adult host. Using this modeling framework, we seek to investigate the stability of the system under a variety of migration dynamics and redistribution assumptions.  We note here that the general system for a two-patch model shown above considers local parasitoid migration between adjacent patches during the vulnerable period.  A global redistribution of hosts and parasitoids is not present in this model, which suggests that the density of hosts and parasitoids at each patch in the next generation is based on the concentration leftover at the end of the previous generation's vulnerable period.  In this sense, the global redistribution into the next generation is completely determined by the continuous dynamics of the previous generation.

\subsection{Constant Migration and Parasitism}\label{Sec21}
We consider the simplest case -- both the migration rates constant, i.e. $g_{ij}(\, \cdot \,) = m_{ij}$.  Because we are investigating the stabilizing effect of migration, we also assume the parasitism rates are constant, i.e. $g_i(\, \cdot \, ) = c_i$.  We note that constant parasitism in this form is analogous to Nicholson-Bailey dynamics at each location \cite{Nicholson_1935, Singh_2007}.  Using these expressions for the migration and parasitism rates, we solve Equations \eqref{dL1_dt} -- \eqref{dP2_dt} subject to the initial conditions in Equations \eqref{ic_start} -- \eqref{ic_end} for all variables (Appendix \ref{App_Constant_No_Redist}).  In Equations \eqref{H} and \eqref{P}, we obtain the following discrete yearly update for the number of hosts and parasitoids at each patch:
\begin{align}
H_{1,t+1} & = RH_{1,t} f_1(\, \cdot \,) \label{H1_con} \\
P_{1,t+1}& = kRH_{1,t}\big[1-f_1(\, \cdot \,)\big] \label{P1_con}\\
H_{2,t+1} & = RH_{2,t} f_2(\, \cdot \,) \label{H2_con} \\
P_{2,t+1}& = kRH_{2,t}\big[1-f_2(\, \cdot \,)\big], \label{P2_con}\end{align}

\noindent where $f_i(\, \cdot \,)$ is the fraction of hosts surviving into the next year at location $i$.  These functions are given by 
\begin{align}
f_1(P_{1,t}, P_{2,t}) & =\exp\left\{ -c_1\left[\frac{m_{21}}{m}(T -\gamma) + \gamma\right] P_{1,t}\right\}\exp\left[ -c_1 \frac{m_{21}}{m}(T-\gamma)P_{2,t}\right] \\ 
f_2(P_{1,t}, P_{2,t}) & =\exp\left\{ -c_2\left[\frac{m_{12}}{m}(T -\gamma) + \gamma\right] P_{2,t}\right\}\exp\left[ -c_2 \frac{m_{12}}{m}(T-\gamma)P_{1,t}\right], \end{align}
where $m = m_{12} + m_{21}$ and $\gamma = (1-e^{-mT})/m$.  The fixed points of this system are 

\begin{align}
H_1^*&  =  \frac{P_1^*}{k(R-1)}\label{H_red1_fix}\\
P_1^* & = \frac{\big[(c_2m_{12}-c_1m_{21})(T-\gamma) + c_2m\gamma\big]\ln(R)}{c_1c_2Tm\gamma} \label{P_red1_fix} \\
H_2^*&  =  \frac{P_2^*}{k(R-1)}\label{H_red2_fix} \\ 
P_2^* & = \frac{\big[(c_1m_{21}-c_2m_{12})(T-\gamma) + c_1m\gamma\big]\ln(R)}{c_1c_2Tm\gamma}. \label{P_red1_fix} \end{align}

\noindent We find that the spectral radius of the Jacobian matrix for the system of four equations given by \eqref{H1_con} -- \eqref{P2_con} is always greater than one for all migration parameter values and for $R>1$ (Appendix \ref{App_Constant_No_Redist}).  This indicates that constant, local migration alone does not provide a stabilizing effect in the host-parasitoid interaction.  This confirms that local migration among two sites is unstable.  

\subsection{Density Dependent Migration}\label{Sec22}
In this section, we investigate the behavior of the system when migration is density dependent.  There are several factors that may influence the decision of a single parasitoid to stay or leave a particular patch.  The presence of substances (i.e. kairomones) secreted by the host informs the parasitoid about patch characteristics such as patch size and host concentration \cite{van_Alphen_2003}.  Thus, rate of migration is inversely proportional to host density.  To model this characteristic, we may define the migration rate from patch $i$ to patch $j$ as the following function: 
\begin{equation} 
g_{ij}(\,\cdot \,)= \frac{m_{ij}}{1 + a L_i(\tau,t)}, \label{L_dependence}\end{equation}
where $a$ is associated to parasitoid searching time similar to that in the model by Huang et.~al.~\cite{Huang_2016}, where the mobile-handling parameter is set to zero, i.e. there is no migration while handling.  Using numerical evidence based on trajectories, we find that dependence on the unparasitized host larvae alone does not yield persistence in the host-parasitoid interaction.  

Furthermore, parasitoids are able to detect if a host is already parasitized by probing it with its ovipositor.  In this case, the host is determined to be unsuitable and if a threshold number of infected hosts are encountered, the parasitoid will leave the patch \cite{van_Alphen_2003}.  In some species, the decision to leave a patch is determined by how many times the parasitoid has oviposited in that patch.  Therefore, to avoid encountering its own oviposited host for a second time, the parasitoid may abandon a patch for other suitable hosts in an adjacent patch \cite{Driessen_1999}.  We assume that the migration rate is directly proportional to the concentration of parasitoids infected host larvae.  We define the migration rate from patch $i$ to patch $j$ as the following function: 
\begin{equation} 
g_{ij}(\,\cdot \,)= m_{ij} I_i(\tau, t).\label{I_dependence}\end{equation}
By a numerical investigation, the model experiences diverging oscillations for both hosts and parasitoids indicating that dependence on the infected host larvae does not induce stability.  

Finally, parasitoids may have incentive to leave a patch based on the presence other parasitoids \cite{Janssen_1995}, in a competitive sense.  We assume the rate of migration to and from patches is linearly dependent on the population of parasitoids at each patch, 
\begin{equation}
g_{ij}(\, \cdot \,) = m_{ij} P_i(\tau, t)\label{P_dependence}\end{equation}
Also, we assume the migration parameters are not equal, i.e. $m_{12} \neq m_{21}$.  Under these conditions, we solve Equations \eqref{dL1_dt} -- \eqref{dP2_dt} subject to the initial conditions in \eqref{ic_start} -- \eqref{ic_end} (Appendix \ref{App_Dependent_No_Redist}).  As in the constant migration case, this yields a discrete yearly update for hosts and parasitoids at each location given by Equations \eqref{H1_con} -- \eqref{P2_con}, with the fraction of hosts surviving at each location given by 

\begin{align}
f_1(P_{1,t}, P_{2,t}) & = \exp\left[ \left( \frac{\sqrt{m_{12}m_{21}}}{|m_{21}-m_{12}|} - \frac{m_{21}}{m_{21} - m_{12}} \right) c_1T P_t\right] \left[ \frac{1 + A_1 e^{\mu T P_t }}{1 + A_1}\right]^{\frac{c_1}{m_{21} - m_{12}}}\\  
f_2(P_{1,t}, P_{2,t}) & = \exp\left[ \left( \frac{\sqrt{m_{12}m_{21}}}{|m_{12}-m_{21}|} - \frac{m_{12}}{m_{12} - m_{21}} \right) c_2T P_t\right] \left[ \frac{1 + A_2 e^{-\mu T P_t }}{1 + A_2}\right]^{\frac{c_2}{m_{12} - m_{21}}}\end{align}
where $P_t = P_{1,t} + P_{2,t}$, $\mu = 2\text{sgn}(m_{12} - m_{21})\sqrt{m_{12}m_{21}}$, and 
\begin{equation*}
A_1  = \frac{\left( \frac{\sqrt{m_{12}m_{21}} }{|m_{21} - m_{12}|} - \frac{m_{21}}{m_{21} - m_{12}} \right) P_t + P_{1,t} }{\left( \frac{\sqrt{m_{12}m_{21}} }{|m_{21} - m_{12}|} + \frac{m_{21}}{m_{21} - m_{12}} \right) P_t -P_{1,t} } , \qquad \quad 
A_2  = \frac{\left( \frac{\sqrt{m_{12}m_{21}} }{|m_{12} - m_{21}|} - \frac{m_{12}}{m_{12} - m_{21}} \right) P_t + P_{2,t} }{\left( \frac{\sqrt{m_{12}m_{21}} }{|m_{12} - m_{21}|} + \frac{m_{12}}{m_{12} - m_{21}} \right) P_t -P_{2,t} }. 
\end{equation*}

\begin{figure}
\begin{center}
\includegraphics[width = \textwidth]{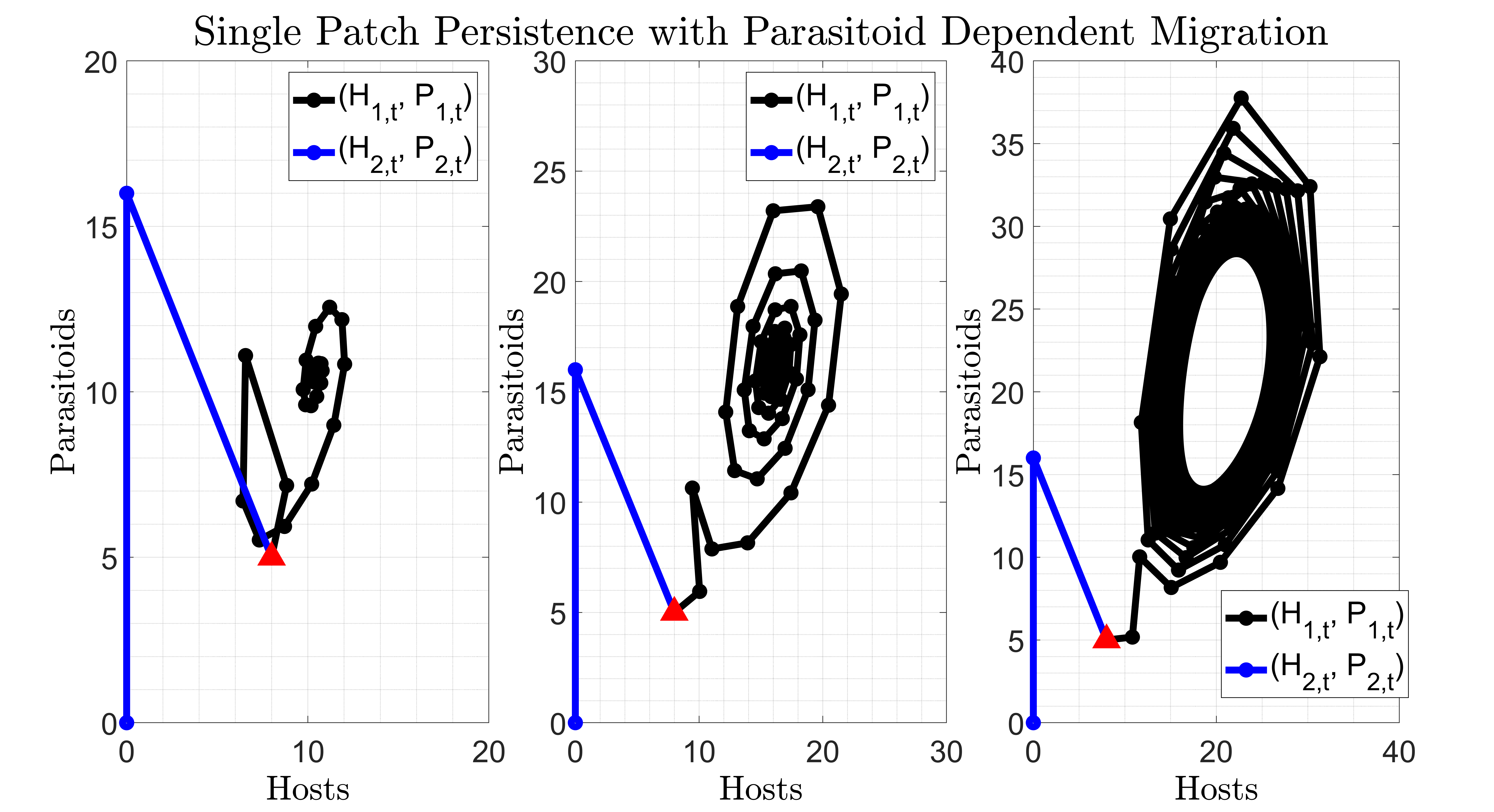}
\end{center}
\caption{\textbf{Persistence is possible at a single location, only if migration rates are asymmetric.}  As the migration rate from patch 1 to patch 2 increases in magnitude, the surviving populations at patch 1 exhibit a stable limit cycle.  Each figure shows trajectories in the phase plane for $(H_{1,t}, P_{1,t})$ and $(H_{2,t}, P_{2,t})$ with $R = 2$,  $c_1 = c_2 = 1$, $T  = 1$, $k = 1$, $m_{21} = 0.01$ and (left) $m_{12} = 8$, (middle) $m_{12} = 12$, and (right) $m_{12} = 16$ with $H_{1,0} = H_{2,0} = 8$, $P_{1,0} = P_{2,0} = 5$. }
\label{No_Redist_Quadratic_Trajs}
\end{figure}

The fixed points for this system cannot be obtained explicitly.  Preliminary trajectories of this system suggest that coexistence is impossible; however, it is possible for only one location to sustain a host and parasitoid population while the other dies out.  Figure \ref{No_Redist_Quadratic_Trajs} shows trajectories in the phase plane for the host and parasitoid populations at each patch for three values of migration rates, where $m_{12} > m_{21}$. The host population at patch 2 dies out in every simulation, but the other population survives as long as the migration rate is not too rapid.  In the case when $m_{12} \gg m_{21}$, the surviving populations will exhibit diverging oscillations.  However, in every case, stability is impossible when the migration rates are comparable in size, $m_{12} \approx m_{21}$.  Overall, this suggests that coexistence can occur at a single patch only if most parasitoids are exiting that patch throughout the vulnerable period.

%

\section{Redistribution}\label{Sec3}
In this section, we consider the effects of global redistribution between each generation coupled with local migration during the vulnerable period.  We assume a dispersion of host larvae and/or adult female parasitoids occurs at the beginning of every season, resulting in a similar redistribution to each patch.  Further, we assume that this dispersion results in an identical proportion to each patch every year.  This is a similar assumption to models considered by Adler \cite{Adler_1993} and Lett et.~al.~\cite{Lett_2003}. To test this assumption on the stability of the system, we modify the initial conditions of the continuous model.  If we only allow the parasitoid population to disperse yearly, the results remain unstable with diverging oscillations.  In the following sections, we discuss the results of redistributing the hosts and redistributing both populations as these yield coexistence. 

\subsection{Redistribution of Hosts}\label{Sec31}

\begin{figure}[b!]
\centering
\includegraphics[width = \textwidth]{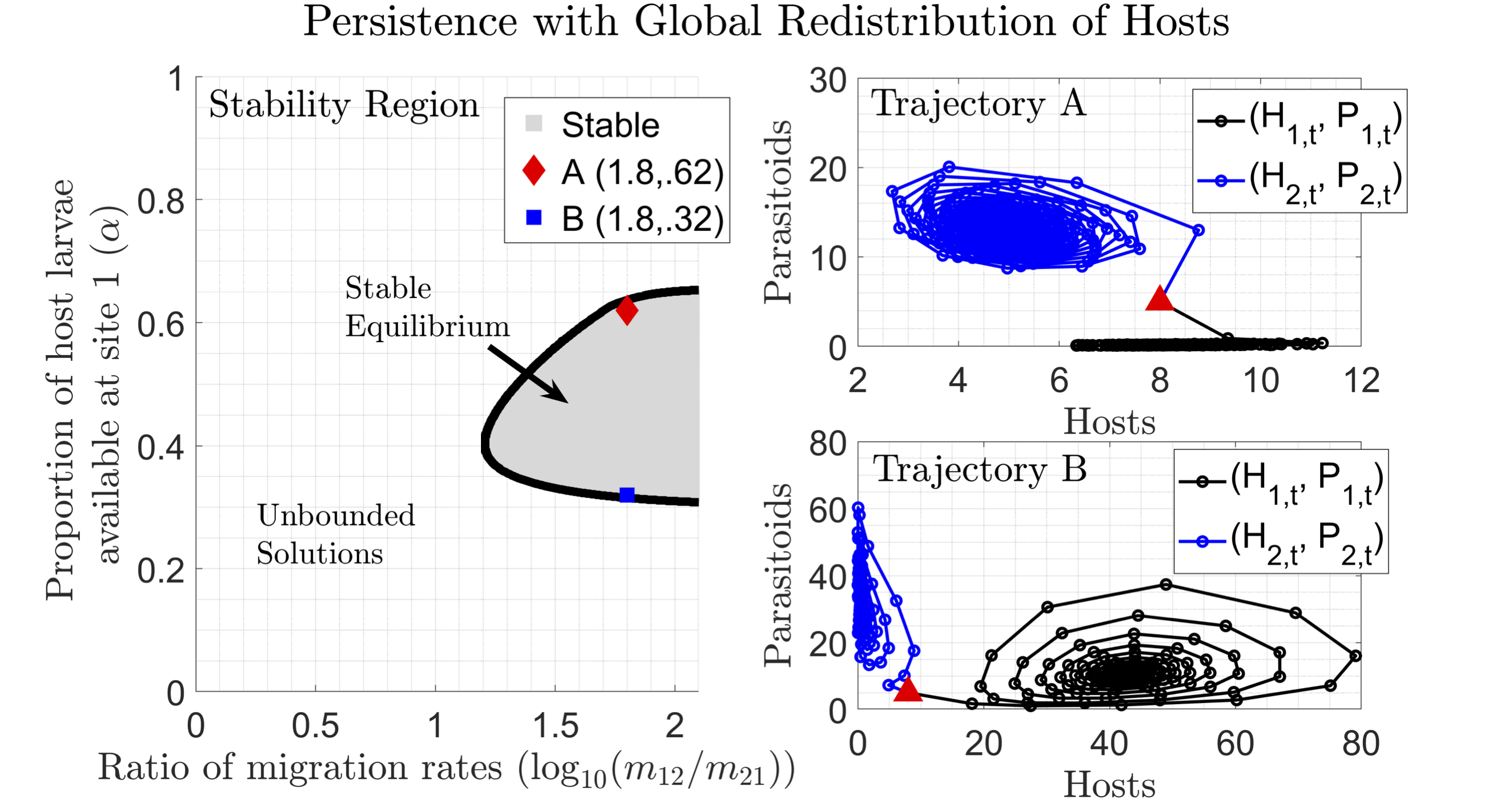}
\caption{\textbf{Global host redistribution stabilizes the constant migration model yielding persistence at both locations for asymmetric local migration rates.}  As the redistribution parameter $\alpha$ increases, the coexistence equilibrium essentially switches from patch 2 to patch 1, indicating a host refuge is present for larger $\alpha$ and a parasitoid refuge is present for smaller $\alpha$.  (Left) A numerical depiction of the stability region in $(\log_{10}(m_{12}/m_{21}), \alpha)$ space, with $R = 2$, $c_1 = c_2 = .1$, $T  = 1$, $k = 1$, and $m_{21} = .1$.  Two points are plotted near the boundary of this stable region and the phase plane trajectories, $(H_{1,t}, P_{1,t})$ and $(H_{2,t}, P_{2,t})$, corresponding to these points are given on the right.  For each trajectory, $H_{1,0} = H_{2,0} = 8$, $P_{1,0} = P_{2,0} = 5$, and (top-right) $m_{12}  = 6.31$ and $\alpha = .62$ and (bottom-right) $m_{12} = 6.31$ and $\alpha = .32$.  }
\label{Host_Redist_Compare}
\end{figure}

We assume that the yearly update for the hosts and parasitoids are subject to identical continuous dynamics during the vulnerable period as in Equations \eqref{dL1_dt} -- \eqref{dP2_dt}, with $g_{ij}(\, \cdot \,)  = m_{ij}$.  The proportion of the total host population, $H_t$, dispersed to patch 1 will be denoted by $\alpha$, which is a redistribution parameter between 0 and 1.  To model this, we modify the initial conditions in \eqref{ic_start} and \eqref{L2_ic} to the following: 

\begin{minipage}{.475\textwidth}
\begin{equation}
L_1(0,t)  = \alpha RH_{t} \label{L1_hred_ic} \end{equation}
\end{minipage}
\begin{minipage}{.475\textwidth}
\begin{equation}
 L_2(0,t) = (1-\alpha) R H_{t}. \label{L2_hred_ic} \end{equation}
\end{minipage}\vspace{.55cm}
In this model, we need only consider the total number of hosts, whose yearly update is given by $H_{t+1} = L_1(T, t) + L_2(T, t)$.  This yields a three dimensional discrete system of the form 
\begin{align}
H_{t+1} & = RH_{t}\left[ \alpha f_1(\, \cdot \,) + (1-\alpha) f_2(\, \cdot \,)\right] \label{H1_hred_con} \\
P_{1,t+1} & = \alpha k RH_{t}\left[ 1 -  f_1(\, \cdot \,)\right] \label{P1_hred_con} \\
P_{2,t+1}& = (1-\alpha)kRH_t\left[ 1 -  f_2(\, \cdot \,) \right] , \label{P2_hred_con}\end{align}
where $f_i$ is the fraction of hosts surviving at location $i$.  These functions are given by 
\begin{align}
f_1(P_{1,t}, P_{2,t}) & =\exp\left\{ -c_1\left[\frac{m_{21}}{m}(T -\gamma) + \gamma\right] P_{1,t}\right\}\exp\left[ -c_1 \frac{m_{21}}{m}(T-\gamma)P_{2,t}\right] \\ 
f_2(P_{1,t}, P_{2,t}) & =\exp\left\{ -c_2\left[\frac{m_{12}}{m}(T -\gamma) + \gamma\right] P_{2,t}\right\}\exp\left[ -c_2 \frac{m_{12}}{m}(T-\gamma)P_{1,t}\right].\end{align}
Numerical simulations of this model show that coexistence is possible.  To better understand the persistence of each population, we numerically compute a stability region in Figure \ref{Host_Redist_Compare}.  Here, we vary the redistribution parameter $\alpha$ and the ratio of migration rates, $\log(m_{12}/m_{21})$.  With $m_{21} = 0.1$, we see that persistence of of hosts and parasitoids only occurs when $m_{12}$ is at least 1.2 orders of magnitude larger than $m_{12}$.  This suggests that coexistence is impossible when the migration between patches is similar, i.e. asymmetric rates yield stability.  Also in Figure \ref{Host_Redist_Compare} is a plot of two stable trajectories corresponding to $\alpha = .62$ and $\alpha = .32$ with asymmetric migration rates.  The parameters associated to these trajectories correspond to points near the boundary of the stability region.  We see that $\alpha = .62$, patch 2 yields persistence of both host and parasitoid populations but patch 1 only consists of a host population as $t\to\infty$.  This suggests that a host refuge is present at the first location as long as the migration to the patch with hosts (i.e. patch 2) is relatively large enough.  In contrast, when $\alpha = .32$, we see the opposite steady-state behavior.  Hosts and parasitoids both persist at patch 1 even though the migration rate away from this patch is much higher.  Patch 2, however, yields only parasitoids as $t\to \infty$.  Thus, depending on the value of $\alpha$, two types of limiting behavior my exist at each location.

\subsection{Redistribution of Hosts and Parasitoids}\label{Sec33}
To obtain analytical results, we assume that both host larvae and adult female parasitoids are subject to a redistribution to each location every year.  To model this, we define the total hosts and total parasitoids in year $t$ as $H_t$ and $P_t$, respectively.  We employ the initial conditions given by Equations \eqref{L1_hred_ic} and \eqref{L2_hred_ic} for the host population and we assume the parasitoid population is redistributed yearly at each location.  The proportion of the total parasitoid population, $P_t$, at patch one will be denoted by $\beta$, a redistribution parameter between 0 and 1.  The initial conditions for the redistribution of parasitoids are given as 

\begin{minipage}{.475\textwidth}
\begin{equation}
P_1(0,t)  = \beta P_{t} \label{P1_pred_ic} \end{equation}
\end{minipage}
\begin{minipage}{.475\textwidth}
\begin{equation}
P_2(0,t) = (1-\beta) P_{t}. \label{P2_pred_ic} \end{equation}
\end{minipage}\vspace{.55cm}

In this model, the update for the total populations is given by
\begin{align}
H_{t+1} & = L_1(T,t) + L_2(T,t) \label{H_poo}\\
P_{t+1}& = kI_1(T,t) + kI_2(T,t). \label{P_poo} \end{align}
We solve this system for all variables in Equations \eqref{dL1_dt} -- \eqref{dP2_dt} (Appendix \ref{App_Constant_Redist}).  We obtain the following discrete yearly update for the number of hosts and parasitoids:
\begin{align}
H_{t+1} & = RH_t f(\, \cdot \,) \label{H_red_con} \\
P_{t+1}& = kRH_t\big[1-f(\, \cdot \,)\big], \label{P_red_con}\end{align}
where $f(\, \cdot \,)$ is the fraction of hosts surviving parasitism at both locations into the next year.  This function is 
\begin{equation} f(P_t) = \alpha e^{-zc_1P_t} + (1-\alpha)e^{-(T-z)c_2P_t} , \label{f_con}\end{equation}
where the parameter $z$ is given by 
\begin{equation}
z = \frac{m_{21}}{m}T + \gamma \left( \beta - \frac{m_{21}}{m} \right). \label{migration_z}\end{equation}

\noindent where $m = m_{12} + m_{21}$ and $\gamma = (1-e^{-mT})/m$.  We note that the parameter $z$ contains all information  about the locatl migration dynamics -- $\beta$ is the proportion of parasitoids starting at the first location, and $m_{12}$ and $m_{21}$ are the migration rates between each location.  We note that $z\in [0,1]$ for all $\beta$, $m_{12}$, and $m_{21}$.  An in depth description of the parameter $z$ is contained in Appendix \ref{App_Constant_Redist}.  In short, $z$ is a measure of the strength of migration from site 2 to site 1.  As we'll see, persistence is only possible when $z\approx 0$.  We wish to do a stability analysis in the remaining parameters of the system, namely $z$, $R$, and $\alpha$, with the parasitism rates,  $c_1$ and $c_2$, held fixed as well as $k$ and $T$.  In this sense, we'll be able to determine the effect of the migration parameter on coexistence of the two species.

\subsubsection{Reduction to Nicholson-Bailey Model}\label{Sec331}
The fraction of hosts surviving parasitism, Equation \eqref{f_con}, is similar to the Nicholson-Bailey model in the sense that $f(P_t)$ is a weighted average of exponential functions, the weight being determined by the proportion, $\alpha$, of hosts at the first location.  Several scenarios of this model will simplify to the unstable Nicholson-Bailey model.  If the entire population of host larvae are at either location at the start of the vulnerable period each year, then the model reduces to the Nicholson-Bailey model.  That is, if $\alpha = 0$ or $\alpha = 1$, then the model is unstable.  When $\alpha = 0$, the fraction of hosts surviving is $f(P_t) = \exp(-zc_1P_t)$.  Similarly, if $\alpha = 1$, then $f(P_t) = \exp[-(1-z)c_2P_t]$, which is essentially identical to the Nicholson-Bailey model.  Also, if the migration parameters, $\beta$, $m_{12}$, and $m_{21}$ are chosen so that $z = 0.5$ and $c_1 = c_2 = c$ (i.e. the parasitism rates are the same at each site), then $f(P_t) = \exp(-cP_t/2)$, which is identical to the Nicholson-Bailey model.  In each scenario, both populations experience diverging oscillations.

\subsubsection{No-Return to Patch 1 ($z = 0$)}\label{Sec332}
We consider a sub-case of the previous model when the migration parameters $\beta$, $m_{12}$, and $m_{21}$ are chosen so that $z = 0$. For instance, if we consider no-return from site 2 to site 1, i.e.~$m_{21} = 0$, and the migration rate from site 1 to site 2 is very fast, i.e.~$m_{12} \to \infty$, then $z \to 0$.  In this case, we see that the parasitoids essentially transport instantaneously to the second location without returning, which results in the entire parasitoid population attacking host larvae at the second location.  The total parasitoid population is trapped at the second location with only a proportion of the host larvae population to oviposit.  We'll see that this is enough to yield coexistence of both species overall.  Fixing the parasitism rates to be equal at each site ($c_1 = c_2 = c$), the stability region is dependent on the parameters $R$ and $\alpha$, which determine the amount of host larvae at each location.  We explore this model below. 

Setting $z = 0$, the fraction of hosts surviving is 
\begin{equation} f(P_t) = \alpha+ (1-\alpha)e^{-cP_t} , \label{f_0}\end{equation}
We find the fixed points of the system in Equations \eqref{H_red_con} and \eqref{P_red_con} as 
\begin{equation}
P_0^*  = \frac{1}{c} \ln\left( \frac{(1-\alpha)R}{1- \alpha R}\right), \qquad \quad H_0^*  = \frac{P_0^*}{k(R-1)}. \label{HP_star_0} \end{equation}
We note that these fixed point values are only valid if $\alpha <1/R$.  A linear stability analysis in Appendix \ref{App_Constant_Redist} shows that coexistence is possible with asymptotically stable solutions occurring when 
\[ \alpha^* < \alpha < \frac{1}{R},\]
where $\alpha^*$ satisfies the following equation:
\[ \frac{(1-\alpha^*R)R}{R-1} \ln\left( \frac{(1-\alpha^*)R}{1-\alpha^* R}\right) = 1.\]

\begin{figure}[b!]
\centering
\includegraphics[width = \textwidth]{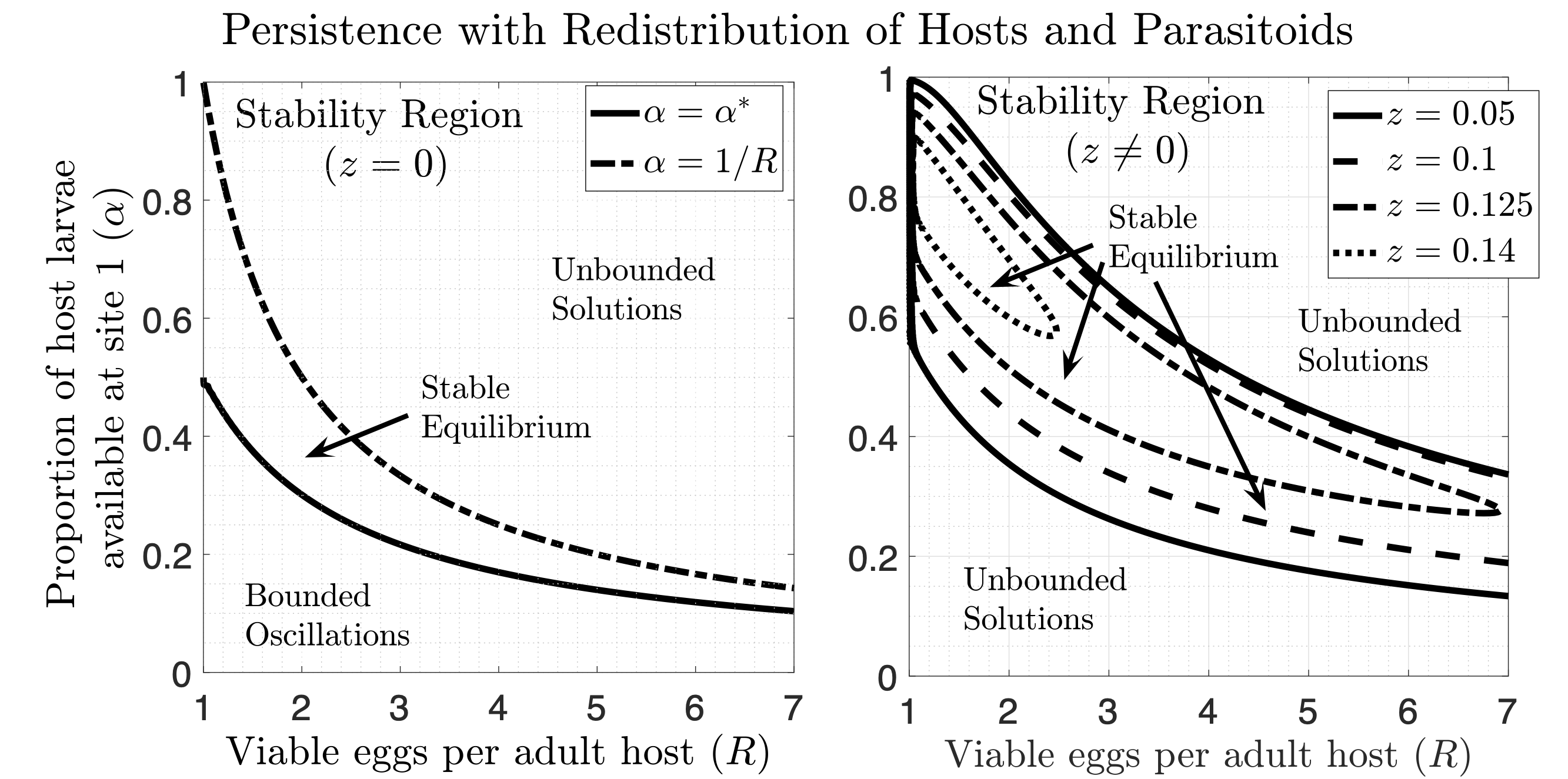}
\caption{\textbf{Global redistribution of hosts and parasitoids stabilizes the constant migration model yielding coexistence when $z = 0$ (no-return) and $z\neq 0$ (asymmetric migration or slow symmetric migration).}  (Left) The stability region in $(R,\alpha)$ space for $z = 0$.  We see that this model yields three stability regimes: bounded oscillations, asymptotically stable, and unbounded oscillations.  The fixed point of the system is asymptotically stable for all $\alpha$ such that $\alpha^* < \alpha < 1/R$.  (Right) The stability region in $(R,\alpha)$ space for $z = 0.05,$ $0.1,$ $0.125,$ and $0.14$.  Within the stable region, coexistence occurs; otherwise, the populations experience diverging oscillations.  }
\label{Stability_Region_both}
\end{figure}

For $\alpha <\alpha^*$, both species experience bounded oscillations.  Figure \ref{Stability_Region_both} depicts the stability region.  When the migration strength from site 2 to site 1 is non-existent ($z = 0$), we see that the proportion of hosts available at site 1 must decrease with $R$ for coexistence to occur.  Essentially, this means that if the parasitoids are trapped at the second location, then there must be a larger proportion of host larvae available for oviposition at the second location every year as the number of viable eggs increases.  However, this region of stability shrinks with $R$, which means the system is less likely to exhibit coexistence if the number of viable eggs is large.  In contrast, we see that as $R\to 1$, the proportion of host larvae available at site 1 need only be more than half for stable conditions to occur.  This means that a host refuge exists at site 1 with the majority of the hosts residing there.  

\subsubsection{Asymmetric migration or slow symmetric migration ($z\neq 0$)}\label{Sec333}
We consider the fixed points of the discrete dynamical system given in Equations \eqref{H_red_con} and \eqref{P_red_con}.  We wish to determine the effect of parasitoid migration on the stability of this system. To this end, we fix the parasitism rates at each location to be $c_1 = c_2 = 1$ with $k =1$ and $T = 1$.  Also, we choose several values for the migration parameter $z$ and determine a stability region in $(R, \alpha)$ for each $z$ value, where $z\in (0,0.5)$.  We find that the fixed points of the system in Equations \eqref{H_red_con} and \eqref{P_red_con}, denoted by $(H^*, P^*)$, for the parasitoid and host larvae population, respectively, satisfy the following equations:
\begin{equation}
f(P^*)  = \frac{1}{R}, \qquad\quad  H^*   = \frac{P^*}{R-1}. \label{fp_1} \end{equation}
The solution for the parasitoid fixed point in Equation \eqref{fp_1} cannot be obtained explicitly.  However, we can numerically compute a stability region by considering the Jury condition $\det[J(H^*,P^*)] = 1$, where $J$ is the Jacobian matrix given in Equation \eqref{J_con}.  

We establish a stability region in $(R,\alpha)$ space for different values of the migration parameter $z$ (Appendix \ref{App_Constant_Redist}).  Figure \ref{Stability_Region_both} gives the stability region for $z = 0.05$, $z = 0.1$, $z = 0.125$, and $z = 0.14$.  We can see that as $z$ increases, the stability region tends to shrink.  This indicates that stability occurs in the presence of relatively unbalanced migration rates as $z$ tends to be much smaller when $m_{12} \gg m_{21}$.  In fact, in the presence of relatively balanced initial parasitoid concentrations (i.e. $\beta \approx 0.5$), $m_{12}$ needs to be approximately two orders of magnitude larger than $m_{21}$ (see Figure \ref{Migration_Parameter_z}) to maintain a migration parameter value of $z = 0.05$.  In this case, a regime of coexistence is observed for $\alpha >0.5$ as long as the number of viable eggs per host ($R$) is relatively small. This indicates that migration from patch 1 to patch 2, with essentially no return to patch 1, creates enough inefficiency in the searching habits of the parasitoid to yield a coexistence between the two populations.  In short, the parasitoids are initially balanced at each patch, but they are quickly leaving patch 1 for patch 2, when most of the host larvae reside at patch 1 (i.e. $\alpha >0.5$).  Also, this indicates coexistence is possible when parasitoids abandon the majority of hosts at site 1 in favor of site 2, creating a host refuge.  Ultimately the decision to leave the patch and barely return causes a stable system.  However, as $R$ increases, the stability region begins to favor values of $\alpha$ less than 0.5, which indicates that reproduction of viable hosts must be relatively high to maintain the majority of parasitoids that are searching at site 2.  As $z$ approaches a threshold value near $0.14$, the region shrinks to a set completely contained above the line $\alpha = 0.5$, indicating that movement away from the patch with the majority hosts is the only condition for coexistence.  Overall, we conclude that for a uniform distribution of parasitoids among sites, the migration rates must be unbalanced in favor of migration to the site with less hosts in order for coexistence to occur. 

\begin{figure}[b!]
\centering
\includegraphics[width = \textwidth]{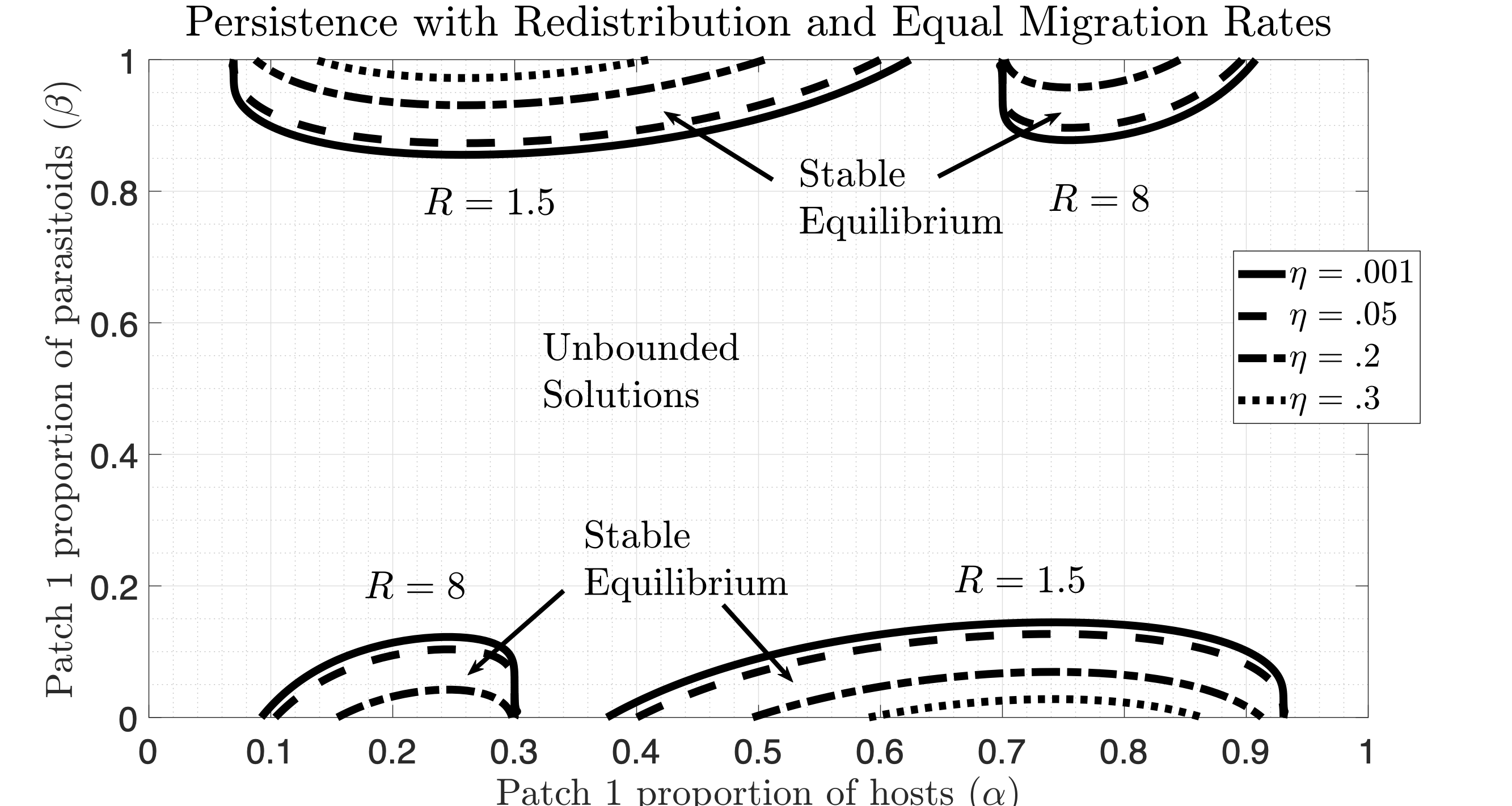}
\caption{\textbf{Coexistence is possible when both populations redistribute yearly with equal local migration rates; however, this rate must be slow.}  As $R$ increases, coexistence only occurs for smaller values of $\eta$, and for relatively equal redistribution proportions.  The stability region is shown for the redistribution model in $(\alpha, \beta)$ space for equal migration rates, $m_{12} = \eta = m_{21}$, where $\eta$ varies from $0.001$ to $0.3$.  Two sets of stability regions are shown for $R = 1.5$ and $R = 8$.  }
\label{Redist_Constant_SR_R_eta}
\end{figure}

To get a better understanding of the stability criteria for equal migration rates, we fix the value of $R$ and consider a stability region in $(\alpha, \beta)$.  Using a similar approach as in Appendix \ref{App_Constant_Redist}, we may work with Equation \eqref{fp_1} and $\det[J(H^*,P^*)] = 1$ to establish how coexistence depends on the redistribution proportions at patch 1 for hosts, $\alpha$, and parasitoids, $\beta$.  Letting $m_{12} = \eta = m_{21}$ and fixing $R = 1.5$, Figure \ref{Redist_Constant_SR_R_eta} gives the stability region for several values of $\eta$.  Here, a stability region exists for migration rates that are relatively low, $\eta < .3$.  We see that the stability region for all values of $\eta$ exists for $\beta$ between approximately 0.4 and 0.9 and $\alpha$ less than or equal to approximately 0.1.  Coexistence will occur in the inverse scenario as well if the proportion of hosts is greater than approximately 0.9 with the proportion of parasitoids between 0.1 and 0.6.  Therefore, for $R = 1.5$ the two species coexist with equal migration rates but only if they are relatively redistributed with their majorities at different locations at the beginning of the vulnerable period, suggesting that a host refuge is present.  Figure \ref{Redist_Constant_SR_R_eta} also shows stability regions for $R = 8$.  As $R$ increases, we not only note the stability region shrinks in size, but the value of $\alpha$ and $\beta$ tend toward each other, indicating that coexistence occurs if the hosts and parasitoids are redistributed with their majorities at the same location.  This is a sharp contrast to the case when $R\approx 1$; however, as $R$ increases, the migration rates must be very small for any type of stability to occur. In fact, the migration rates are $\sim 10^{-1}$, suggesting that most parasitoids do not migrate during the vulnerable period since $T = 1$.  We note that this model is a generalization of the Lett et.~al.~model \cite{Lett_2003} and the Adler model \cite{Adler_1993}.  Allowing $\eta \to 0$ gives $z \to \beta T$, which would mean there is no local migration during the vulnerable period. This yields identical results to Lett's aggregate model (global redistribution parameters: $\alpha = v_1^*$ and $\beta T c_1 = \mu_1^*$).

\section{Functional Response }\label{Sec4}
In this section, we consider linear dependence on the current host larvae density in the parasitism rate, i.e. $g_i(\, \cdot\,) = c_i L_i$, which incorporates a quadratic functional response.  For a single location, the quadratic functional response model is known to yield neutrally stable oscillations in both the host and parasitoid populations \cite{Singh_2007}.  In the following sections, we incorporate the functional response into the two-patch model and determine whether global and/or local migration can stabilize the system.

\subsection{Functional Response with Constant Migration}\label{Sec41}
We investigate the behavior of the system when a functional response is applied to the parasitism rates with migration held constant.  We assume $g_{ij}(\, \cdot \,) = m_{ij}$ and the parasitism is linearly dependent on the host population at each site, $g_{i}(\,\cdot\,) = c_iL_i$ as in \cite{Singh_2007}.  Under these conditions, we solve Equations \eqref{dL1_dt} -- \eqref{dP2_dt} subject to the initial conditions in \eqref{ic_start} -- \eqref{ic_end} (Appendix \ref{App_Functional_No_Redist}).  As in the constant migration case, this yields an equivalent discrete yearly update for hosts and parasitoids at each location as Equations \eqref{H1_con} -- \eqref{P2_con} but with the fraction of hosts surviving at each location given by 
\begin{align}
f_1(H_{1,t},P_{1,t},P_{2,t}) & = \frac{1}{1 + c_1\left\{ \left[ \frac{m_{21}}{m}(T-\gamma) + \gamma\right]P_{1,t} + \frac{m_{21}}{m}(T-\gamma)P_{2,t}\right\} RH_{1,t}} \label{f1_no_redist_fun} \\ 
f_2(H_{2,t},P_{1,t},P_{2,t}) & = \frac{1}{1 + c_2\left\{ \left[ \frac{m_{12}}{m}(T-\gamma) + \gamma\right]P_{2,t} + \frac{m_{12}}{m}(T-\gamma)P_{1,t}\right\} RH_{2,t}}, \label{f2_no_redist_fun}\end{align}
where $m = m_{12} + m_{21}$ and $\gamma = (1-e^{-mT})/m$.  Similar to the constant parasitism case, constant migration does not have a stabilizing effect on the functional response model without redistribution.  The two patch model remains neutrally stable for all parameter values.

\subsection{Functional Response with Density Dependent Migration}\label{Sec42}
As in Section \ref{Sec22}, we consider three separate cases where the migration rate is density dependent.  Parasitoids typically migrate from patch 1 to patch 2 if the host density at the patch 1 is low.  Also, if the infected host larvae density or the parasitoid density is high at patch 1, a parasitoid may decide to leave patch 1 for patch 2.  To cover these three cases, we let $g_{ij}(\, \cdot\,)$ be defined by Equations \eqref{L_dependence}, \eqref{I_dependence}, or \eqref{P_dependence} coupled with the functional response in parasitism.  Under these conditions, we numerically integrate Equations \eqref{dL1_dt} -- \eqref{dP2_dt} subject to the initial conditions \eqref{ic_start} -- \eqref{ic_end}.  

\begin{figure}[h!]
\begin{center}
\includegraphics[width = \textwidth]{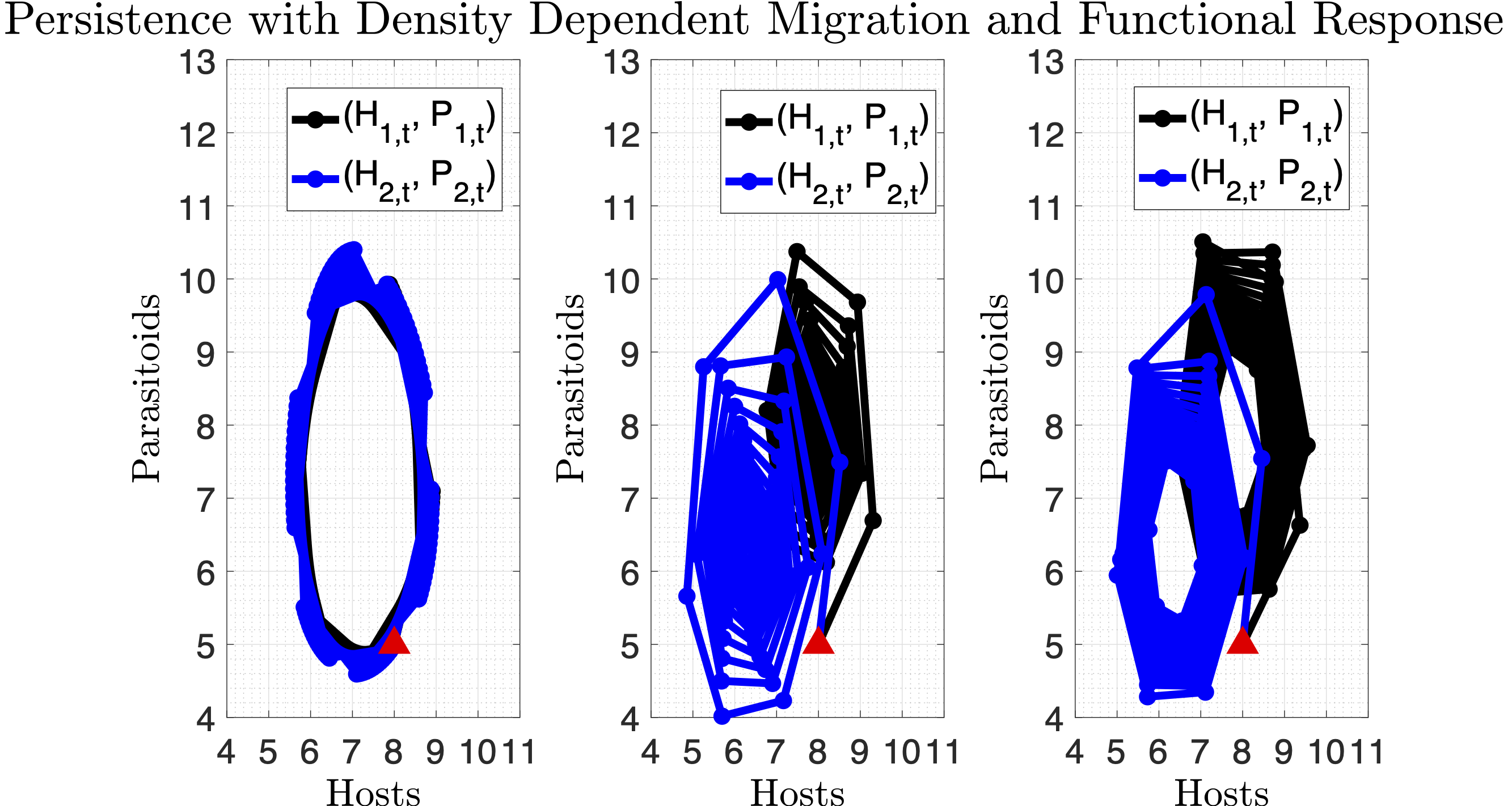}
\end{center}
\caption{\textbf{Persistence is possible when all three types of density dependent migration are coupled with functional response.}  The type of dependence yields different rates of convergence.  Trajectories in the phase plane for $(H_{1,t}, P_{1,t})$ and $(H_{2,t}, P_{2,t})$ with $R = 2$,  $c_1 = c_2 = 0.01$, $T  = 1$, $k = 1$, $m_{21} = 0.1$ and $m_{12} = 0.2$ using three density dependent migration rates: (left) $g_{ij}(\, \cdot \,) = m_{ij}/(1+L_i(t,\tau))$, (middle) $g_{ij}(\,\cdot \,) = m_{ij}I_i(t,\tau)$, and (right) $g_{ij}(\,\cdot\,) = m_{ij}P_i(t,\tau)$ with $H_{1,0} = H_{2,0} = 8$, $P_{1,0} = P_{2,0} = 5$. }
\label{No_Redist_Fun_Quadratic_Trajs}
\end{figure}

A numerical investigation show that the three systems all yield stable, non-zero equilibria at both locations.  In fact, stable trajectories are observed when the migration rates are relatively equal.  Figure \ref{No_Redist_Fun_Quadratic_Trajs} shows an output of the functional response model with equal parameters in all three density dependent cases.  In contrast to the constant migration models of Section \ref{Sec22}, both patches exhibit non-zero population values as $t \to \infty$.  However, the rate of convergence for each model seems to depend on the type of density dependence migration, with the rate of convergence slowest if the migration is inversely proportion to the current host larvae population.

\subsection{Functional Response with Redistribution of Hosts}\label{Sec43}
\begin{figure}[b!]
\centering
\includegraphics[width = \textwidth]{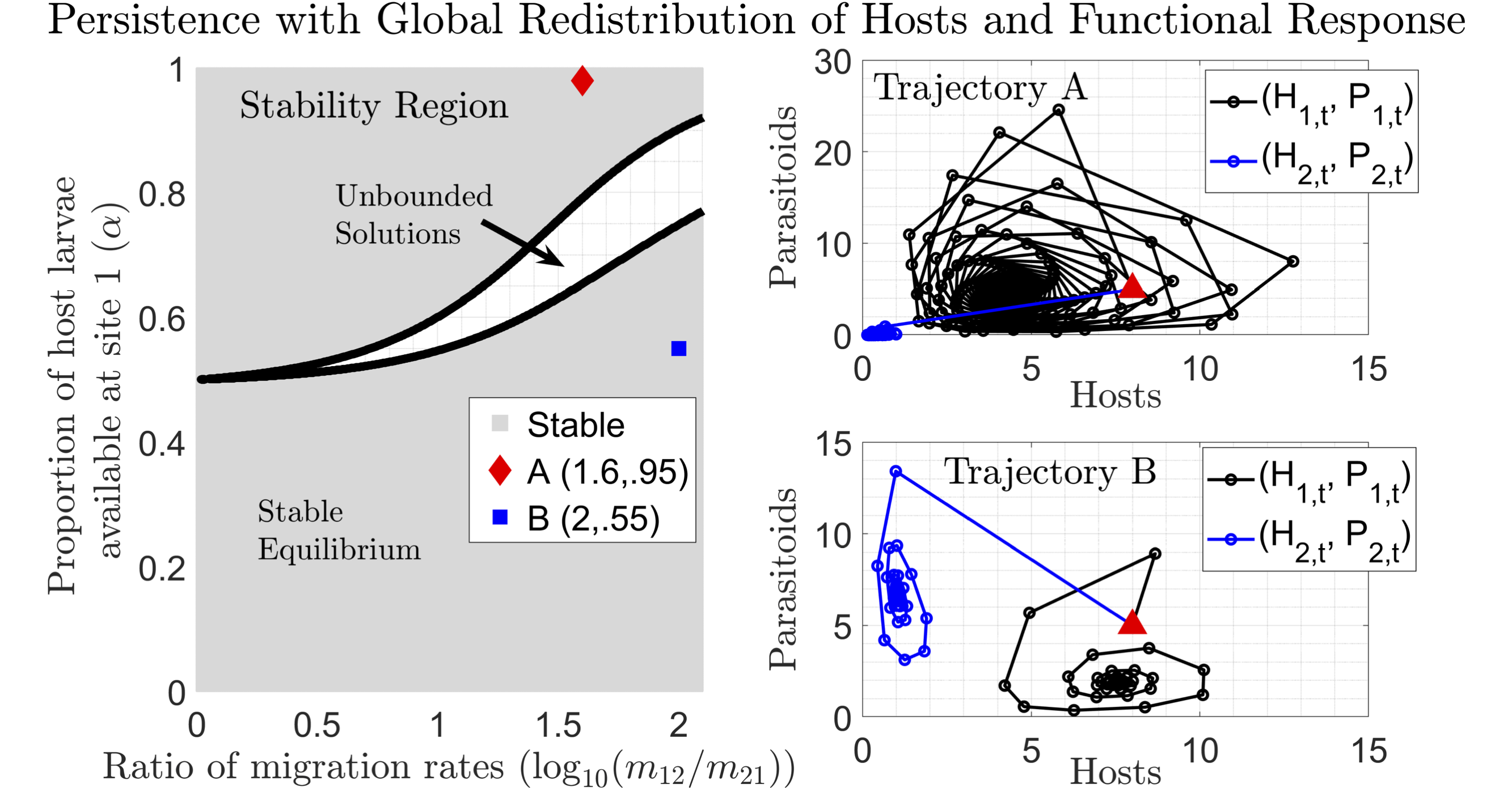}
\caption{\textbf{Functional response further stabilizes the global host redistribution model with constant migration yielding persistence at both locations with symmetric and asymmetric migration rates.}  (Left) A numerical depiction of the stability region in $(\log_{10}(m_{12}/m_{21}), \alpha)$ space, with $R = 2$, $c_1 = c_2 = .1$, $T  = 1$, $k = 1$, and $m_{21} = .1$.  Two points are plotted near the boundary of this stable region and the phase plane trajectories, $(H_{1,t}, P_{1,t})$ and $(H_{2,t}, P_{2,t})$, corresponding to these points are given on the right.  For each trajectory, $H_{1,0} = H_{2,0} = 8$, $P_{1,0} = P_{2,0} = 5$, and (top-right) $m_{12}  = 3.98$ and $\alpha = .95$ and (bottom-right) $m_{12} = 10$ and $\alpha = .55$.  }
\label{Host_Redist_Functional_Compare}
\end{figure}

We consider the functional response model with the host population redistribution as explained in Section \ref{Sec31}.  This yields a three dimensional discrete system equivalent to Equations \eqref{H1_hred_con} -- \eqref{P2_hred_con} with the fraction of hosts surviving at each location given by 
\begin{align}
f_1(H_t,P_{1,t}, P_{2,t}) & = \frac{1}{1 + \alpha c_1\left\{ \left[ \frac{m_{21}}{m}(T-\gamma) + \gamma\right]P_{1,t} + \frac{m_{21}}{m}(T-\gamma)P_{2,t}\right\} RH_t}  \label{f1_hred_fun}\\ 
f_2(H_t,P_{1,t}, P_{2,t}) & =\frac{1}{1 + (1-\alpha)c_2\left\{ \left[ \frac{m_{12}}{m}(T-\gamma) + \gamma\right]P_{2,t} + \frac{m_{12}}{m}(T-\gamma)P_{1,t}\right\} RH_t}.\label{f2_hred_fun}\end{align}
As in the constant parasitism case, the redistribution of hosts has a stabilizing effect to the original neutrally stable model.  Figure \ref{Host_Redist_Functional_Compare} shows a numerically constructed stability region according to the redistribution parameter $\alpha$ and the ratio of migration rates.  Compared to Figure \ref{Host_Redist_Compare}, we see that there is a much larger region of parameter space that yields a stable equilibrium.  In fact, the functional response model is much more stable when the migration rates are relatively equal in magnitude.  Interestingly, however, a region of instability is possible if the migration rates are asymmetric in magnitude with $\alpha >0.5$.  This suggests that local migration during the vulnerable period has a destabilizing effect when asymmetric redistribution of hosts is present.  The figure confirms that the model is neutrally stable if $\alpha = 0.5$ with equal migration rates.  For two points in the parameter space, we also show the trajectories at each location.  Most trajectories contained in the stability region will yield non-zero equilibria at both locations, but we can see that for large $\alpha$, the population at patch 2 seems to approach zero in Trajectory A.  This suggests, as in the case with constant migration, that various types of equilibrium behavior is present within the stability region.

%

\subsection{Functional Response with Redistribution of Hosts and Parasitoids}\label{Sec44}
We consider a redistribution of both populations as explained in Section \ref{Sec33} with a functional response applied to the parasitism at each location.  We solve the system for all variables in Equations \eqref{dL1_dt} -- \eqref{dP2_dt} subject to the redistribution initial conditions given by Equations \eqref{L1_hred_ic} and \eqref{L2_hred_ic} for the host population and Equations \eqref{P1_pred_ic} and \eqref{P2_pred_ic} for the parasitoid population. (Appendix \ref{App_Functional_Redist}).  This yields the following update equivalent to Equations \eqref{H_red_con} and \eqref{P_red_con} with the fraction of hosts surviving to the next year given by 
\begin{equation} 
f(H_t, P_t) = \frac{\alpha }{1 + c_1 z \alpha R H_t P_t} + \frac{1-\alpha }{1 + c_2(T-z)(1- \alpha) R H_t P_t}. \label{f_fun} \end{equation}
We let $T = 1$ for the remainder of the section.  The lumped parameter $z = z(\beta,m_{12},m_{21})$ also appears in this model and is given by Equation \eqref{migration_z}.  As in Section \ref{Sec31}, we wish to do a stability analysis in the parameters $z$, $R$, and $\alpha$, with the functional response rates of parasitism, $c_1$ and $c_2$ fixed.  We will fix the values of $k$ and $T$ as well as these parameters are known to not influence stability.

\subsubsection{Reduction to Neutrally Stable Model}\label{Sec441}
The fraction of hosts surviving parasitism, Equation \eqref{f_fun}, is similar to the neutrally stable model considered by Singh et.~al.~in \cite{Singh_2007} in the sense that $f(H_t, P_t)$ is a weighted average of reciprocal functions.  Here, the weight is determined by the yearly redistribution of hosts at site 1, given by $\alpha$.  This model will yield identical results to the quadratic functional response model in \cite{Singh_2007}, if $\alpha = 0$ or $\alpha = 1$.  In either of these cases, all hosts are at a single location subject to a functional response in the parasitism rate.  Also, the model is neutrally stable if $\alpha + z = 1$ and the rates of parasitism are equivalent ($c_1 = c_2 = c$).  In this case, the fraction of hosts surviving is exactly 
\begin{equation}
f(H_t, P_t) =  \frac{1}{1 + cz\alpha RH_t P_t}, \end{equation} 
which has the same form as the Singh et.~al.~quadratic functional response model with an extra factor of $z\alpha $ in the denominator.  This model will yield neutrally stable trajectories with a period of $2\pi/\arctan(\sqrt{R^2 - 1})$ for small amplitude oscillations.  The neutral stability is not dependent on the value of $c$, $z$, or $\alpha$.

\subsubsection{No-Return to Patch 1 ($z = 0$)}\label{Sec442}
As in the model with constant parasitism and migration, we assume the lumped parameter takes on a value of $z = 0$.  This reduces the model considerably and gives insight into what happens when the migration rate from patch 1 to patch 2 is very fast or if the migration rate from patch 2 to patch 1 is essentially zero. Fixing $c_1 = c_2 = c$ and $z = 0$, the fraction of hosts surviving to the next year is 
\begin{equation}
f(H_t, P_t) = \alpha + \frac{1-\alpha}{1 + c(1-\alpha)RH_tP_t}, \label{f0_fun} \end{equation}
which reveals the following equilibrium point 
\begin{align}
H_0^* & = \frac{P_0^*}{k(R-1)} \label{H0_fun} \\
P_0^* & = (R-1)\sqrt{\frac{k}{c(1-\alpha)(1-\alpha R)R}}. \label{P0_fun} \end{align}
A linear stability analysis in Appendix \ref{App_Functional_Redist} shows that $\det\big[J(H_0^*, P_0^*)\big] < 1$ for all $\alpha < 1/R$ and $R>1$.  Therefore, the model is asymptotically stable for all values of $\alpha$ and $R$ for which it exists.  A similar analysis holds for $z = 1$.

\subsubsection{Asymmetric migration or slow symmetric migration ($z\neq 0$)}\label{Sec442}
We consider the fixed point of the discrete dynamical system given in Equations \eqref{H_red_con} and \eqref{P_red_con} with the fraction of hosts surviving given by Equation \eqref{f_fun}, where $z \in (0, 0.5)$.  The fixed points of this satisfy the following equations: 
\begin{align}
\frac{H^*}{P^*} & = \frac{1}{k(R-1)} \label{H_star_fun}\\ 
H^*P^* & = -\frac{ (1-\alpha R)k_2 + \big(1-(1-\alpha)R\big)k_1 }{2k_1k_2R} +\cdots \notag \\ 
& \qquad \frac{ \sqrt{\left[ (1-\alpha R)k_2 + \big(1-(1-\alpha)R\big)k_1\right]^2 - 4k_1k_2(R-1)}}{2k_1k_2R}, \label{P_star_fun} \end{align}
where $k_1 = c_1\alpha z$ and $k_2 = c_2(1-z)(1-\alpha)$.  A brief stability analysis in Appendix \ref{App_Functional_Redist} shows that this fixed point is asymptotically stable for all $\alpha$, $z$, and $R$ except when $\alpha + z = 1$; in this case, the fixed point is neutrally stable.

\section{Discussion}\label{Sec5}
In this paper, we have considered the classic host-parasitoid interaction coupled with global redistribution of both populations and local migration of female parasitoids.  In contrast to most phenomenological models, we incorporate the migration dynamic into the semi-discrete framework, which tends to have more relevance  to parasitoid populations with one year life cycles.  This approach also allows us to track the local migration tendencies of parasitoids during the vulnerable period.  The preceding analyses show that constant local migration between two patches cannot stabilize Nicholson-Bailey parasitism at each site.  In fact, local migration coupled with a neutrally stable functional response will also not stabilize host-parasitoid interactions.  However, if local migration is density dependent, a stable equilibrium is formed.  Further, coupling location migration with global redistribution has a quantifiable stabilizing effect in host-parasitoid interactions, which is known to be the case in other studies.  A summary of all stabilizing factors are contained in Table \ref{Summary_Table}.  Overall, we find that the asymmetric local migration rates typically stabilize the two-patch semi-discrete model; however, in the case of global redistribution of hosts only or both hosts and parasitoids, persistence can occur with relatively equal local migration rates. 

\begin{table}[b!]
\centering
\begin{adjustbox}{max width=\textwidth}
\setlength{\aboverulesep}{0pt}
\setlength{\belowrulesep}{0pt}
\label{Summary_Table}
\begin{tabular}{| c | c | c | c |}
\hline
Parasitism & Local Migration & Global Redistribution & Stability \\\hline\hline
\multirow{8}{*}{\makecell{Constant\\ $g_{i}(\,\cdot\,) = c_{i}$}}
    & \multirow{6}{*}{\makecell{Constant \\ $g_{ij}(\, \cdot \,) = m_{ij}$}}
        & None  & Unstable    \\
        \cmidrule{3-4}
    	&   & Hosts only 
		& \makecell{Coexistence/ Host Refuge possible \\ (asymmetric, see Fig.~\ref{Host_Redist_Compare})}\\
	\cmidrule{3-4}
	&  & Hosts and Parasitoids 
		& \makecell{Coexistence/ Host Refuge possible \\ (asymmetric or slow \\ symmetric, see Fig.~\ref{Stability_Region_both})} \\ 
      
    \cmidrule{2-4}
    & \makecell{Parasitoid Dependent \\ $g_{ij}(\, \cdot \,) = m_{ij}P_i$}
    	& None
	    &  \makecell{Single Patch Coexistence \\  (asymmetric)}  \\
    \hline\hline
\multirow{9}{*}{\makecell{Functional \\ Response \\ $g_i(\, \cdot \,) = c_i L_i$}}
    & \multirow{5}{*}{\makecell{Constant \\ $g_{ij}(\, \cdot \,) = m_{ij}$}}
        & None  & Neutrally Stable    \\
        \cmidrule{3-4}
    	&   & Hosts only 
		& \makecell{Coexistence \\ (see Fig.~\ref{Host_Redist_Functional_Compare})} \\
	\cmidrule{3-4}
	&  & Hosts and Parasitoids 
		& \makecell{Stable for $\alpha + z \neq 1$\\ Neutrally Stable otherwise.} \\     
	\cmidrule{2-4}
 & \makecell{Density Dependent \\ $g_{ij}(\, \cdot \,) = m_{ij}/(1+L_i)$\\  $g_{ij}(\, \cdot \,) = m_{ij}I_i$\\  $g_{ij}(\, \cdot \,) = m_{ij}P_i$}
    	& None
	    & Always Stable  \\\hline

\end{tabular}
\end{adjustbox}

\caption{\textbf{Summary of migration tendencies that stabilize the two-patch semi-discrete model.}  Persistence may occur for asymmetric local migration rates, i.e.~$m_{12} \gg m_{21}$, or slow symmetric, i.e.~$m_{12}\approx m_{21}\approx 10^{-1}$.  Global redistribution coupled with constant local migration and density dependent local migration stabilize the two-patch semi-discrete model under both constant and functional response parasitism.    }
\end{table}

We found, using numerical evidence, that density dependent local migration is stabilizing in both the constant and functional response parasitism models.  We considered three types of density dependent migration: inversely proportional to the host larvae population, directly proportional to the infected host population, directly proportional to parasitoid population. In the case of constant parasitism, coexistence is only observed at the a single patch for asymmetric migration rates.  Indeed, persistence at patch 1 occurs when $g_{12} = m_{12}P_1$ with $m_{12}$ at least three orders of magnitude greater than $m_{21}$.  With $T = 1$, this essentially means all parasitoids migrate to patch 2 and never return.  Hence, stability is caused by the parasitoids leaving the patch that has a host population.  In the case of a quadratic functional response in the attack rates, Singh and Nisbet show that the sole equilibrium point is neutrally stable \cite{Singh_2007}. In the two-patch semi-discrete model, we find that any type of density dependence in the migration rate yields stability.  In fact, the migration rates can be relatively equal, suggesting that there is a dynamic exchange of parasitoids between the patches during the vulnerable period.  

Redistribution of the host population can stabilize the two-patch model.  This means the host population is reset to similar proportions every year.  In the case of constant parasitism, stability can arise as long as $m_{12}$ is at least one order of magnitude larger than $m_{21}$.  Furthermore, the proportion of hosts at patch 1 ($\alpha$) must be contained between approximately $0.35$ and $0.65$.  For $\alpha$ closer to $0.65$ and $log_{10}(m_{12}/m_{21}) = 1.8$, coexistence occurs in the form of a host refuge.  We find that coexistence occurs at patch 2, where there are less hosts every year; however, the parasitoids are traveling to patch 2 more frequently and so little to no parasitoids reside at patch 1, leaving the host population to steadily oscillate there.  This suggests that persistence is defined by parasitoids leaving the patch that has the larger proportion of hosts, which is a reoccurring theme in the stability criteria. If $log_{10}(m_{12}/m_{21}) = 1.8$ and $\alpha$ is closer to $0.35$, we see the opposite: a coexistence equilibrium exists at patch 1 with a parasitoid population existing at patch 2 (with very little to no hosts).  This suggests that the system is stable when parasitoids are moving to the patch with a larger host population.  The coexistence occurs where the lesser of the two populations reside, which is completely opposite of the previous case.  We see that this scenario leads to the parasitoids decimating the host population at site 2, yet still surviving from year-to-year.  In the case of functional response in the parasitism rate, we see a much larger region of stability.  In fact, for equal migration rates, the model is stable for all values of $\alpha$, except for $\alpha = 0.5$, suggesting that there is a dynamic exchange of parasitoids between both patches during the vulnerable period. When $\alpha = 0.5$, the model is identical to the single patch model by Singh and Nisbet \cite{Singh_2007}.  As $m_{12}$ becomes larger than $m_{21}$, a region of instability arises for $\alpha > 0.5$, suggesting that persistence is impossible if more parasitoids are located at the site with less hosts. 

Global redistribution of both populations stabilizes the two-patch semi-discrete model.  In this case, both populations reset at each locations with similar proportions every year.  As in the previous models, an offset ratio of local migration rates will yield stability in the system.  If we consider relatively equal migration rates, $m_{12}\approx m_{21}$, a value of $z \approx 0.01$ (i.e.~the strength of migration from site 2 to site 1 is weak) is maintained when $\beta \approx 0$.  However, in this situation both migration rates are of order $10^{-2}$.  This means that coexistence will occur within the system if there is negligible patch-use with the majority of the parasitoids searching patch 2, the location with minimal hosts under the condition that $R$, the number of viable eggs per adult host,  is approximately less than 4.  As $R$ increases, we note again that stability occurs for values of $\alpha$ less than $0.5$, indicating that more hosts are at patch 2 than patch 1.  Therefore, in the case of increasing $R$ and negligible yet relatively equal migration rates, as long as the majority of parasitoids coincide with the majority of hosts at the same patch at the beginning of the vulnerable period every year, coexistence can occur.  This is a heavy contrast to the case when $R \to 1$ as the parasitoids and hosts must be, essentially, in different places to yield coexistence.  As $z$ increases towards $0.14$, coexistence is maintained for migration rates between $10^{-1}$ and $10^0$, but we see that the stability region shrinks to a set bounded below by $\alpha = 0.5$.  Hence, coexistence is only maintained if the parasitoids are located at the site with less hosts, but have a tendency to search both sites at equal pace.  



\appendix
\section{ODE Solutions and Stability Analysis to Constant Parasitism Model with Constant Migration}\label{App_Constant_No_Redist}

We seek the solutions to the ODE system given by Equations \eqref{dL1_dt} -- \eqref{dP2_dt} subject to the initial conditions in Equations \eqref{ic_start} -- \eqref{ic_end}.  We assume the parasitism rates are constant, $g_1(\,\cdot\,) = c_1$ and $g_2(\,\cdot\,) = c_2$.  Also, the migration rates are constant, $g_{12}(\,\cdot\,) = m_{12}$ and $g_{21}(\,\cdot\,) = m_{21}$.  Let $m = m_{12} + m_{21}$ and $P_t = P_{1,t} + P_{2,t}$.  Then the solutions for the parasitoid populations at each location are given by 
\begin{align}
P_1(\tau, t) & = \frac{m_{21}}{m} P_t + \left( P_{1,t} - \frac{m_{21}}{m} P_t\right) e^{-m\tau} \label{p1_cc} \\ 
P_2(\tau, t) & = \frac{m_{12}}{m} P_t + \left( P_{2,t} - \frac{m_{12}}{m} P_t\right) e^{-m\tau}, \label{p2_cc}\end{align}
and the solutions for the host populations are 
\begin{align}
L_1(\tau, t) & = R H_{1,t} \exp\left\{-c_1\left[\frac{m_{21}}{m} P_t \tau + \left( P_{1,t} - \frac{m_{21}}{m}P_t\right)\left(\frac{1-e^{-m\tau}}{m}\right)\right]\right\}  \\ 
L_2(\tau, t) & = R H_{2,t} \exp\left\{-c_2\left[\frac{m_{12}}{m} P_t \tau + \left( P_{2,t} - \frac{m_{12}}{m}P_t\right)\left(\frac{1-e^{-m\tau}}{m}\right)\right]\right\}. \end{align}
It follows immediately from the initial conditions that $I_1(\tau, t) = RH_{1,t} - L_1(\tau, t)$ and $I_2(\tau, t) = RH_{2,t} - L_2(\tau, t)$.

We consider the stability of the fixed points given by Equations \eqref{H_red1_fix} -- \eqref{P_red1_fix}.  Noting that $f_i(P_1^*, P_2^*) = 1/R$, for $i = 1, 2$, we may write the Jacobian matrix evaluated at the fixed point of the system of four equations as the block matrix
\begin{equation} J  = \begin{bmatrix} I & \frac{1}{k}A \\ k(R-1)I & -A \end{bmatrix}, \end{equation}
where $I$ is the $2\times 2$ identity matrix and
\begin{equation} A = \frac{R}{R-1} \left. \begin{bmatrix} P_1\pd{f_1}{P_1} & P_1 \pd{f_1}{P_2}  \\ 
  P_2\pd{f_2}{P_1} & P_2\pd{f_2}{P_2} \end{bmatrix}\right|_{P_1 = P_1^*, P_2 = P_2^*}. \end{equation}
Interestingly, we find that the determinant of $J$ is given by $\det(J) = R^2\det(A)$ and the Eigenvalues of $J$ are roots of the following characteristic polynomial
\begin{equation} \lambda^2(\lambda-1)^2 + \lambda (\lambda - 1) (\lambda - R) \text{tr}(A) + (\lambda - R)^2 \det(A) = 0. \end{equation}
Let $\mu_{1}$ and $\mu_2$ be the eigenvalues of $A$, then the eigenvalues of $J$ are given by
\begin{align*}
\lambda_{1,2}  & = \frac{1-\mu_1}{2} \pm \sqrt{(1-\mu_1)^2 + 4R\mu_1} \\ 
\lambda_{3,4}  & = \frac{1-\mu_2}{2} \pm \sqrt{(1-\mu_2)^2 + 4R\mu_2}. \end{align*}
We find that the spectral radius of $J$ is always greater than one for all migration parameter values, indicating that the nontrivial fixed point given by Equations \eqref{H_red1_fix} -- \eqref{P_red1_fix} is asymptotically unstable for $R>1$.

\section{ODE Solutions to the Constant Parasitism Model with Parasitoid Dependent Migration}\label{App_Dependent_No_Redist}
We seek the solutions to the ODE system given by Equations \eqref{dL1_dt} -- \eqref{dP2_dt} subject to the initial conditions in Equations \eqref{ic_start} -- \eqref{ic_end}.  We assume the parasitism rates are constant, $g_1 = c_1$ and $g_2 = c_2$.  Also, the migration rates are linearly dependent on the parasitoid population at each location, $g_{12}(\,\cdot\,) = m_{12}P_1$ and $g_{21}(\,\cdot\,) = m_{21}P_2$.  Let $m = m_{12} + m_{21}$ and $P_t = P_{1,t} + P_{2,t}$, then the solutions for the parasitoid populations at each location are given by 
\begin{align}
P_1(\tau, t) & = \frac{m_{21}}{m_{21}-m_{12}} P_t - \frac{\sqrt{m_{12}m_{21}}}{|m_{21} - m_{12}|} \left[ \frac{1 - A_1 e^{\mu P_t \tau} }{1 + A_1 e^{\mu P_t \tau}} \right] P_t \\ 
P_2(\tau, t) & = \frac{m_{12}}{m_{12}-m_{21}} P_t - \frac{\sqrt{m_{12}m_{21}}}{|m_{12} - m_{21}|} \left[ \frac{1 - A_2 e^{-\mu P_t \tau} }{1 + A_2 e^{-\mu P_t \tau}} \right] P_t \\ 
\end{align}
where $P_t = P_{1,t} + P_{2,t}$, $\mu = 2\text{sgn}(m_{12} - m_{21})\sqrt{m_{12}m_{21}}$ and 
\begin{equation*}
A_1  = \frac{\left( \frac{\sqrt{m_{12}m_{21}} }{|m_{21} - m_{12}|} - \frac{m_{21}}{m_{21} - m_{12}} \right) P_t + P_{1,t} }{\left( \frac{\sqrt{m_{12}m_{21}} }{|m_{21} - m_{12}|} + \frac{m_{21}}{m_{21} - m_{12}} \right) P_t -P_{1,t} } , \qquad \quad 
A_2  = \frac{\left( \frac{\sqrt{m_{12}m_{21}} }{|m_{12} - m_{21}|} - \frac{m_{12}}{m_{12} - m_{21}} \right) P_t + P_{2,t} }{\left( \frac{\sqrt{m_{12}m_{21}} }{|m_{12} - m_{21}|} + \frac{m_{12}}{m_{12} - m_{21}} \right) P_t -P_{2,t} } 
\end{equation*}
The solutions for the host populations are 
\begin{align}
L_1(\tau, t) & = R H_{1,t} \exp\left[ \left( \frac{\sqrt{m_{12}m_{21}}}{|m_{21}-m_{12}|} - \frac{m_{21}}{m_{21} - m_{12}} \right) c_1 P_t \tau\right] \left[ \frac{1 + A_1 e^{\mu P_t \tau}}{1 + A_1}\right]^{\frac{c_1}{m_{21} - m_{12}}}\\ 
L_2(\tau, t) & = R H_{2,t} \exp\left[ \left( \frac{\sqrt{m_{12}m_{21}}}{|m_{12}-m_{21}|} - \frac{m_{12}}{m_{12} - m_{21}} \right) c_2 P_t \tau\right] \left[ \frac{1 + A_2 e^{-\mu P_t \tau}}{1 + A_2}\right]^{\frac{c_2}{m_{12} - m_{21}}}\end{align}
It follows immediately from the initial conditions that $I_1(\tau, t) = RH_{1,t} - L_1(\tau, t)$ and $I_2(\tau, t) = RH_{2,t} - L_2(\tau, t)$.

\section{ODE Solutions and Stability Analysis to the Constant Parasitism Model with Redistribution}\label{App_Constant_Redist}

\begin{figure}[h!]
\centering
\includegraphics[width = \textwidth]{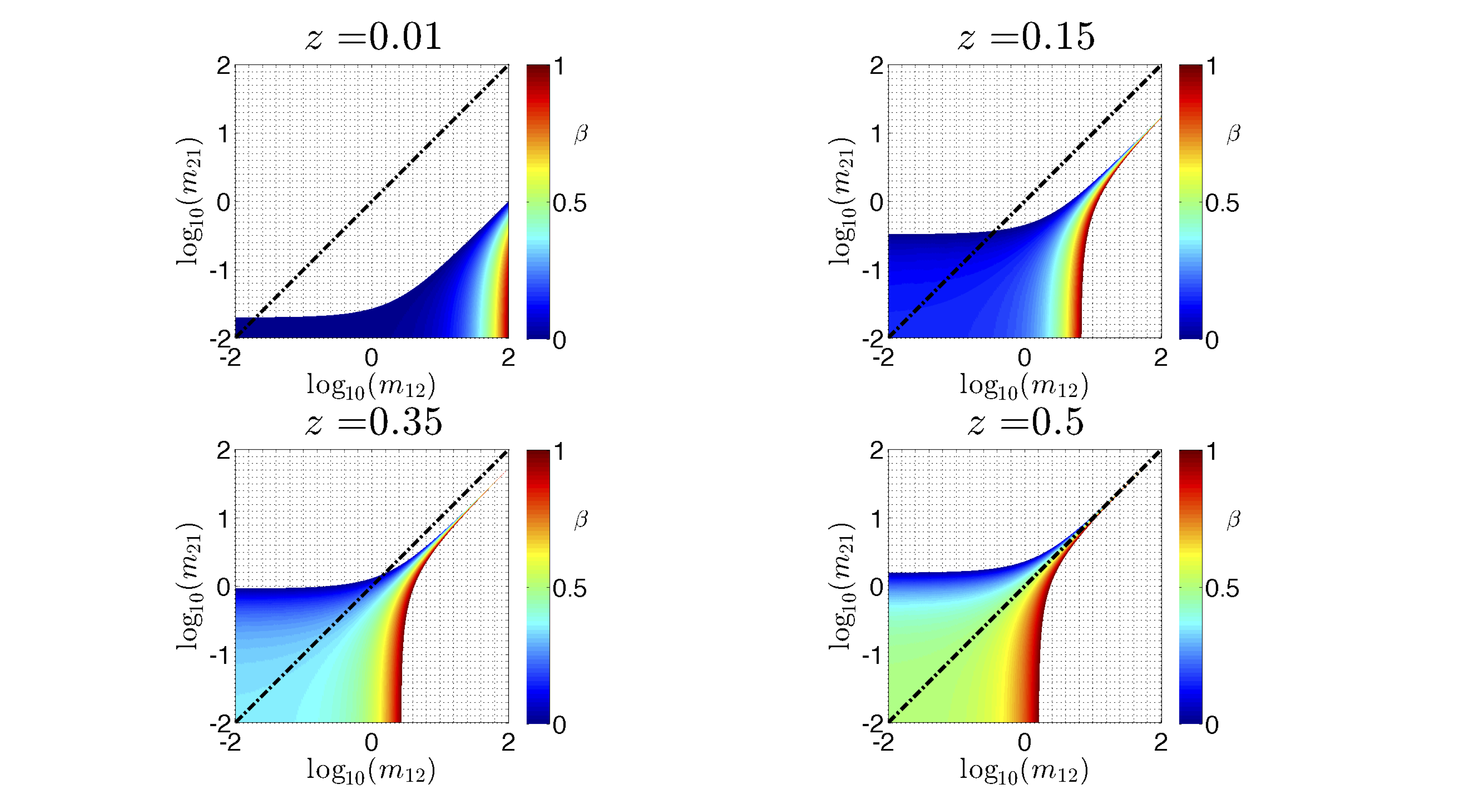}
\caption{Iso-surfaces for the migration parameter $z = z(\beta, m_{12}, m_{21})$ defined by Equation \eqref{migration_z} with $T = 1$.  The surfaces are generated for $z = 0.01$ (top-left), $z = 0.015$ (top-right), $z = 0.35$ (bottom-left), and $z = 0.5$ (bottom-right) and viewed from above, with the colormap corresponding to the value of $\beta$.  We see that to attain a small value of $z$, which represents the strength of migration from site 2 to 1, asymmetric migration rates are needed for relatively equal distribution of parasitoids (i.e. $\beta \approx 0.5$). }
\label{Migration_Parameter_z}
\end{figure}

The solutions for the parasitoid populations at each location to the ODE system given by Equations \eqref{dL1_dt} -- \eqref{dP2_dt} with $g_{1}(\,\cdot\,) = c_1$ and $g_2(\,\cdot\,) = c_2$ subject to the redistribution initial conditions in Equations \eqref{L1_hred_ic} -- \eqref{L2_hred_ic} and Equations \eqref{P1_pred_ic} -- \eqref{P2_pred_ic} are given by 
\begin{align}
P_1(\tau, t) & = \frac{m_{21}}{m} P_t + \left( \beta - \frac{m_{21}}{m} \right) P_t e^{-m\tau} \label{P1_constant}\\
P_2(\tau, t) & = \frac{m_{12}}{m} P_t + \left( 1-\beta - \frac{m_{12}}{m} \right) P_t e^{-m\tau} \label{P2_constant}, \end{align}
where $m = m_{12} + m_{21}$.  The solutions for the host populations are 
\begin{align}
L_1(\tau, t) & = \alpha RH_t \exp\left\{ -c_1\left[ \frac{m_{21}}{m}\tau + \left( \beta - \frac{ m_{21}}{m} \right)\left( \frac{1 - e^{-m\tau}}{m}\right)\right] P_t \right\} \\
L_2(\tau, t) & = (1-\alpha) RH_t \exp\left\{ -c_2\left[ \frac{m_{12}}{m}\tau + \left(1- \beta - \frac{m_{12}}{m} \right)\left( \frac{1 - e^{-m\tau}}{m}\right)\right] P_t \right\} \end{align}
It follows immediately from the initial conditions that $I_1(\tau, t) = \alpha RH_{t} - L_1(\tau, t)$ and $I_2(\tau, t) = (1-\alpha) RH_{t} - L_2(\tau, t)$.  Defining the migration parameter $z$ as 
\begin{equation}
z = \frac{m_{21}}{m}T + \gamma \left( \beta - \frac{m_{21}}{m} \right). \label{migration_z}\end{equation}
gives the fraction of host surviving in Equation \eqref{f_con}.  Letting $T = 1$, we note that the parameter $z = z(\beta, m_{12}, m_{21})$ contains all information about the local migration dynamics -- $\beta$ is the proportion of parasitoids starting at the first location, and $m_{12}$ and $m_{21}$ are the migration rates between each location.  We note that $z\in [0,1]$ for all $\beta$, $m_{12}$, and $m_{21}$.  Here, $z$ is a measure of the strength of migration from site 2 to site 1.  Figure \ref{Migration_Parameter_z} depicts a top down view of iso-surfaces for specific values of $z$.  Considering the top-left figure of Figure \ref{Migration_Parameter_z}, we see that when $z$ is small, i.e. $z = 0.01$, the parameter $m_{12}$ is generally much larger than $m_{21}$ for values of $\beta > 0.5$.  Small values of $z$ can also be obtained if the migration parameters are small with relatively equal magnitudes and the proportion of parasitoids at site 1 is small, i.e. $\beta \approx 0$.  In this case, a weak strength of migration from patch 2 to 1 is attained because almost all parasitoids are at site 2 and the migration rates are negligible but relatively equal.  Considering the bottom-right figure of Figure \ref{Migration_Parameter_z}, we see that $z = 0.5$ corresponds to migration rates being relatively equal for any value of $\beta$.  However, if the migration rates are relatively small but not equal, we see that $z = 0.5$ can be established if the proportion of parasitoids at site 1 balances the strength of movement from site 2 to site 1.  Similar behavior is observed for values of $z$ greater than 0.5 -- the value $1 - z$ is the strength of migration of parasitoids from site 1 to site 2.  In Sections \ref{Sec332} and \ref{Sec333}, we perform a stability analysis in the remaining parameters of the system, namely $z$, $R$, and $\alpha$, with the parasitism rates,  $c_1$ and $c_2$, held fixed as well as $k$ and $T$.  

\textsc{For $z = 0$:}  We consider the stability of the fixed points that satisfy Equation \eqref{HP_star_0}.  
Define the right hand side of Equations \eqref{H_red_con} and \eqref{P_red_con} as the following functions: $F(H,P) = RH f(P)$ and $G(H, P) = kRH\big[1-f(P)\big]$.  We can see that $k\partial F/\partial P = -\partial G/ \partial P$.  We obtain the general form of the Jacobian matrix to the system evaluated at $(H, P)$ as the following: 
\begin{equation}J(H, P) = \displaystyle \begin{bmatrix} 1 & \frac{RPf'(P)}{k(R-1)} \\ k(R-1) &  \frac{RPf'(P)}{1-R} \end{bmatrix}.\label{J_con}\end{equation}
From Equations \eqref{HP_star_0}, we have 
\begin{equation} f'(P_0^*)  = \frac{c(\alpha R-1)}{R}.\end{equation}
Using this expression, we obtain the Jacobian matrix evaluated at $(H_0^*, P_0^*)$ as
\begin{equation} J(H_0^*,P_0^*) = \begin{bmatrix} 1 & \left( \frac{\alpha R - 1}{R-1} \right) \ln \left( \frac{(1-\alpha)R}{1-\alpha R} \right) \\  
R-1 & \left( \frac{1-\alpha R}{R-1} \right) \ln \left( \frac{(1-\alpha)R}{1-\alpha R} \right) \end{bmatrix} .\end{equation} 
We note that the Jacobian doesn't depend on the parasitism rate $c$ or $k$, only the number of viable adult hosts, $R$, and the proportion of initial host larvae at the start of the vulnerable period, $\alpha$. Using the Jury conditions, (namely $\det[J(H_0^*,P_0^*)] < 1$), we create Figure \ref{Stability_Region_both} that demonstrates the stability region of this model.  We see that coexistence is possible if $\alpha < 1/R$, with asymptotically stable solutions occurring when 
\[ \alpha^* < \alpha < \frac{1}{R},\]
where $\alpha^*$ satisfies the following equation:
\[ \frac{(1-\alpha^*R)R}{R-1} \ln\left( \frac{(1-\alpha^*)R}{1-\alpha^* R}\right) = 1,\]
which is equivalent to the boundary of the Jury condition $\det[J(H_0^*,P_0^*)] < 1$.

\begin{figure}[h!]
\centering
\subfloat{%
\resizebox*{6.5cm}{!}{\includegraphics{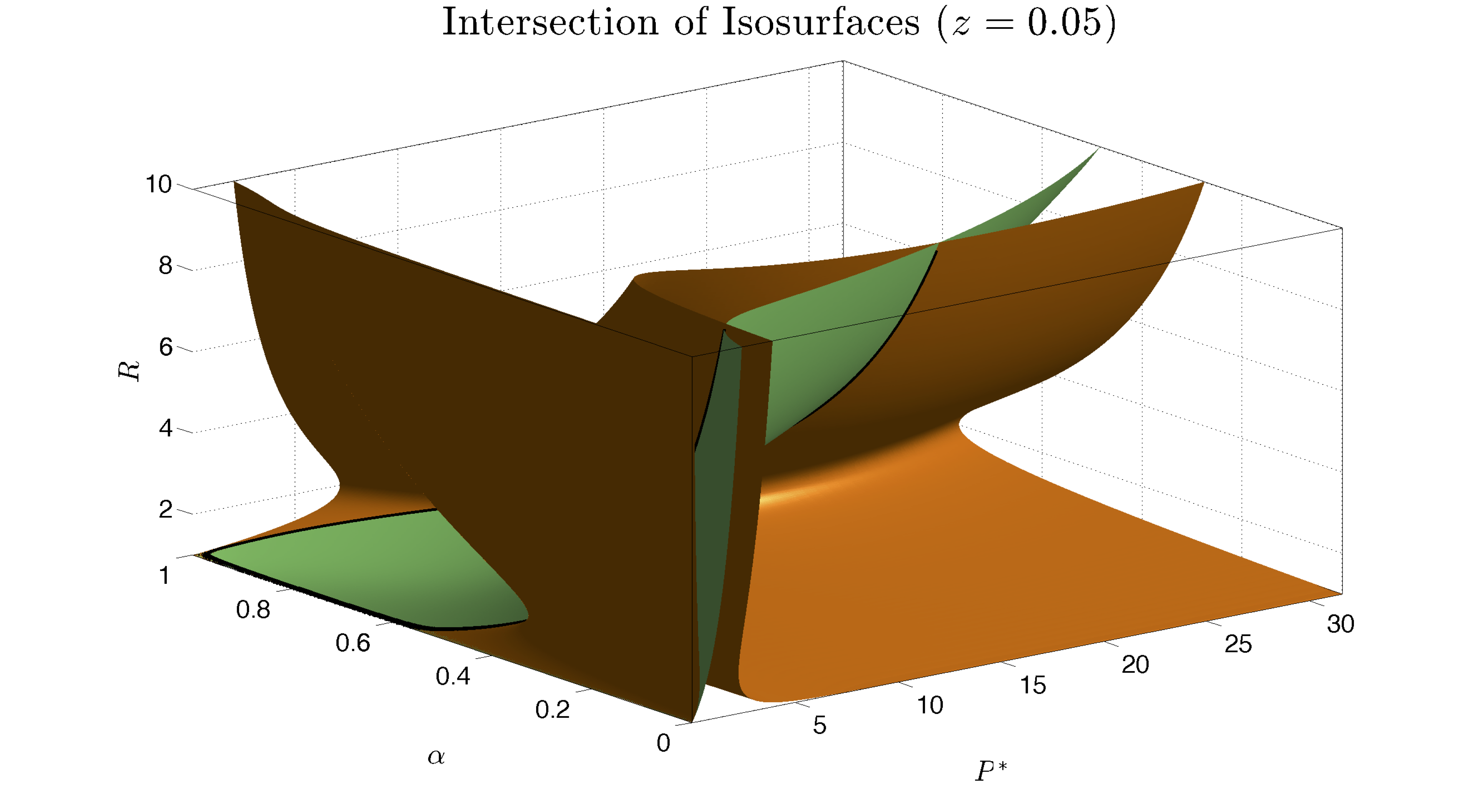}}}\hspace{5pt}
\subfloat{%
\resizebox*{7.5cm}{!}{\includegraphics{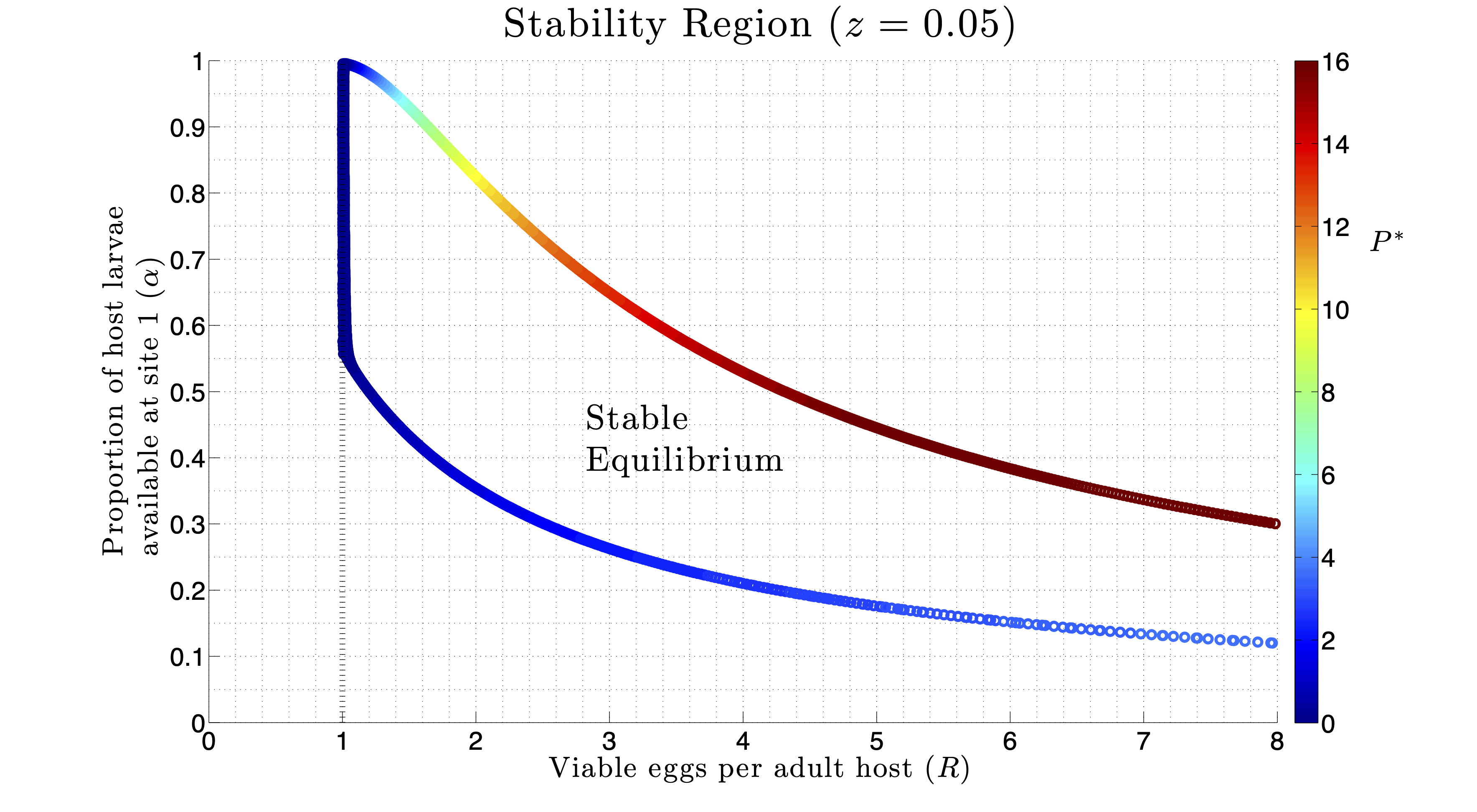}}}
\caption{(Left) The black space curve defines the intersection of the two iso-surfaces in $(P^*, \alpha, R)$ space defined by Equation \eqref{fp_1} (green) and Equation \eqref{Jury_1} (orange). (Right) The space curve that defines the intersection boundary is plotted in the $(R,\alpha)$ plane, with the colormap corresponding to the value of $P^*$.    }
\label{iso-surfaces}
\end{figure}

\textsc{For $z\in(0, 0.5)$:}  We consider the stability of the fixed points that satisfy Equation \eqref{fp_1}.  The solution for the parasitoid fixed point cannot be obtained explicitly.  To determine the boundary of the stability region, we work with Equation \eqref{fp_1} and the Jury condition that determines the stability boundary, namely $\det[J(H^*,P^*)] = 1$, where $J$ is the Jacobian matrix given in Equation \eqref{J_con}.  This yields the equation 
\begin{equation}
\frac{R^2 P^* f'(P^*)}{1-R} = 1, \label{Jury_1}\end{equation}
\noindent where
\begin{equation}
f'(P^*) = -\alpha z e^{-zP^*} - (1-\alpha)(1-z)e^{-(1-z)P^*}. \label{f_prime}\end{equation}
We note that, for fixed $z$, Equations \eqref{fp_1} and \eqref{Jury_1} define two implicit surfaces in $(R,\alpha, P^*)$ space.  To determine the stability region in $(R,\alpha)$ space, we first consider the space curve defined by the intersection of these two surfaces for a fixed value of $z$. The intersection of the two surfaces for $z = 0.05$ is depicted in Figure \ref{iso-surfaces}.  If we view this intersection as a space curve in $(R, \alpha, P^*)$, we can determine the stability region for values of $R$ and $\alpha$ (also Figure \ref{iso-surfaces}).  We establish the various stability regions for different values of $z$ in Figure \ref{Stability_Region_both} by computing the intersection of the iso-surfaces in this way.  

\section{ODE Solutions to the Functional Response Model with Constant Migration}\label{App_Functional_No_Redist}
We consider the solutions to the ODE system given by Equations \eqref{dL1_dt} -- \eqref{dP2_dt} subject to the initial conditions in Equations \eqref{ic_start} -- \eqref{ic_end}.  We assume the parasitism rates are dependent on the host larvae population, $g_1(\,\cdot\,) = c_1L_1$ and $g_2(\,\cdot\,) = c_2L_2$ and the migration rates are constant.  The solutions for the parasitoid populations at each location are identical to those in Equations \eqref{p1_cc} and \eqref{p2_cc}.  The host populations are found to be
\begin{align}
L_1(\tau, t) & = \frac{RH_{1,t} }{1 + c_1 \left[ \frac{m_{21}}{m} P_t \tau + \left( P_{1,t} - \frac{m_{21}}{m}P_t\right)\left(\frac{1-e^{-m\tau}}{m} \right) \right] RH_{1,t}} \\ 
L_2(\tau, t) & = \frac{RH_{2,t} }{1 + c_2 \left[ \frac{m_{12}}{m} P_t \tau + \left( P_{2,t} - \frac{m_{12}}{m}P_t\right)\left(\frac{1-e^{-m\tau}}{m} \right) \right] RH_{2,t}}, \end{align}
where $P_t = P_{1,t} + P_{2,t}$.  It follows immediately from the initial conditions that $I_1(\tau, t) = RH_{1,t} - L_1(\tau, t)$ and $I_2(\tau, t) = RH_{2,t} - L_2(\tau, t)$.

\section{ODE Solutions and Stability Analysis to Functional Response Model with Redistribution}\label{App_Functional_Redist}
The solutions for the parasitoid populations at each location to the ODE system given by Equations \eqref{dL1_dt} -- \eqref{dP2_dt} with $g_{1}(\,\cdot\,) = c_1L_1$ and $g_2(\,\cdot\,) = c_2L_2$ subject to the redistribution initial conditions in Equations \eqref{L1_hred_ic} -- \eqref{L2_hred_ic} and \eqref{P1_pred_ic} -- \eqref{P2_pred_ic} are identical to those provided in Equations \eqref{P1_constant} and \eqref{P2_constant}.  The solutions for the host populations are 
\begin{align}
L_1(\tau, t) & = \frac{\alpha R H_t}{1 + c_1\left[\frac{m_{21}}{m}\tau + \left(\beta - \frac{m_{21}}{m} \right)\left(\frac{1 - e^{-m\tau}}{m}\right)\right] \alpha R H_t P_t}\\
L_2(\tau, t) & = \frac{(1-\alpha) R H_t}{1 + c_2\left[\frac{m_{12}}{m}\tau + \left(1-\beta - \frac{m_{12}}{m} \right)\left(\frac{1 - e^{-m\tau}}{m}\right)\right](1- \alpha) R H_t P_t} \end{align}
It follows immediately from the initial conditions that $I_1(\tau, t) = \alpha RH_{t} - L_1(\tau, t)$ and $I_2(\tau, t) = (1-\alpha) RH_{t} - L_2(\tau, t)$.

\textsc{For $z = 0$:}  We consider the stability of the fixed points given by Equations \eqref{H0_fun} and \eqref{P0_fun}.  We note that these fixed point values are only valid for $\alpha < 1/R$.  Let $F(H, P) = RH_t f(H, P)$ and $G(H, P) = kRH\left[1 - f(H,P)\right]$, where $f(H, P)$ is given by Equation $\eqref{f0_fun}$.  The Jacobian matrix to this system evaluated at $(H,P)$ is given by 
\begin{equation} J(H,P) = \begin{bmatrix} Rf + RH\pd{f}{H} & RH\pd{f}{P} \\ kR(1-f) - kRH\pd{f}{H} & -kRH\pd{f}{P} \end{bmatrix} \label{Jac_fun}\end{equation}
Noting that $f(H_0^*, P_0^*) = 1/R$ and 
\begin{equation} \pd{f}{H}(H_0^*, P_0^*) = - \frac{c(1-\alpha R)^2}{R} P_0^*, \qquad \quad \pd{f}{P}(H_0^*, P_0^*) = -\frac{c(1-\alpha R)^2}{kR(R-1)}P_0^*, \end{equation}
it follows that 
\begin{equation} \det\big[J(H_0^*, P_0^*)\big] = \frac{1-\alpha R}{1-\alpha}.\end{equation}
The Jury condition responsible for the stability of the model, namely $\det\big[J(H_0^*, P_0^*)\big] < 1$, holds for all $\alpha < 1/R$ and $R>1$.  Therefore, the model is asymptotically stable for all values of $\alpha$ and $R$ for which it exists.  A similar analysis holds for $z = 1$.

\textsc{For $z\in(0,0.5)$:}  We consider the stability of the fixed points that satisfy Equations \eqref{H_star_fun} and \eqref{P_star_fun}.  Noting that $f(H^*, P^*) = 1/R$, we find that the determinant of the Jacobian matrix (Equation \eqref{Jac_fun}) evaluated at the point $(H^*, P^*)$ is 
\begin{equation}
\det\big[ J(H^*, P^*) \big] = \frac{R^3H^* P^*}{R-1} \left[ \frac{\alpha k_1}{(1 + k_1 RH^*P^*)^2} + \frac{ (1-\alpha)k_2 }{(1 + k_2 R H^* P^*)^2} \right]. \end{equation}
This determinant is always between 0 and 1 for all values of $\alpha$, $z$, and $R$, except in the case when $\alpha + z = 1$.  In that case, the determinant is equal to one and the fixed point is neutrally stable.  

\bibliographystyle{tfs} 
\bibliography{Parasitoid_Bibliography.bib}

\end{document}